\documentclass[journal]{IEEEtran}
\usepackage{amsmath,amssymb,amsfonts,amsmath}
\usepackage{bm}
\usepackage{algpseudocode}
\usepackage[ruled,linesnumbered]{algorithm2e}
\algdef{SE}{Begin}{End}{\textbf{begin}}{\textbf{end}}
\usepackage{booktabs}
\usepackage{graphicx}
\usepackage{multirow}
\usepackage{array}
\usepackage{cite}
\usepackage{textcomp}
\usepackage{xcolor}
\usepackage{booktabs}

\newcommand{\vect}{\mathbf}
\newcommand{\HT}{\mathsf{H}}

\newcommand{\CBF}{CBF}

\usepackage{amsmath}
\DeclareMathOperator*{\argmin}{arg\,min}
\newcommand{\argmax}{\mathop{\rm arg\,max}\limits}
\usepackage{color}

\newcommand{\IVAconv}{CIVA}
\newcommand{\IVA}{IVA}
\newcommand{\swIVAconv}{swCIVA}
\newcommand{\swIVA}{swIVA}
\newcommand{\swWPE}{swWPE}
\newcommand{\trick}{coarse-fine source model}

\title{
Switching Independent Vector Analysis and \\Its Extension to Blind and Spatially Guided Convolutional Beamforming Algorithms
}

\author{Tomohiro Nakatani,~\IEEEmembership{Fellow,~IEEE,}
        Rintaro Ikeshita,~\IEEEmembership{Member,~IEEE,}
        Keisuke Kinoshita,~\IEEEmembership{Senior Member,~IEEE,}\\
        Hiroshi Sawada,~\IEEEmembership{Fellow,~IEEE,}
        Naoyuki Kamo, 
        Shoko Araki,~\IEEEmembership{Fellow,~IEEE,}
\thanks{T. Nakatani, R. Ikeshita, H. Sawada, K. Kinoshita, N. Kamo, and S. Araki are with NTT Corporation. }
\thanks{Manuscript received July 27, 2021; revised December 2, 2021 and January 27, 2022.}}

\usepackage[absolute,showboxes]{textpos}
\TPMargin{5pt}
\newcommand{\copyrightstatement}{
\begin{textblock}{0.8}(0.1,0.01)
\noindent
\footnotesize
\copyright 2022 IEEE.  Personal use of this material is permitted.  Permission from IEEE must be obtained for all other uses, in any current or future media, including reprinting/republishing this material for advertising or promotional purposes, creating new collective works, for resale or redistribution to servers or lists, or reuse of any copyrighted component of this work in other works.
\end{textblock}
}
\begin{document}
\copyrightstatement

\maketitle
\begin{abstract}
This paper develops a framework that can accurately perform denoising, dereverberation, and source separation using a relatively small number of microphones. It has been empirically confirmed that Independent Vector Analysis ({\IVA}) can blindly separate $N$ sources from their sound mixture even with diffuse noise when a sufficiently large number ($=M$) of microphones are available (i.e., $M\gg N)$. However, the estimation accuracy is seriously degraded when the number of microphones, or more specifically $M-N$ $(\ge 0)$, decreases. To overcome this {\IVA} limitation, we propose switching IVA ({\swIVA}) in this paper. With {\swIVA}, the time frames of an observed signal with time-varying characteristics are clustered into several groups, each of which can be well handled by {\IVA} with a small number of microphones, and thus accurate estimation can be achieved by individually applying {\IVA} to each group. 
Conventionally, a switching mechanism was introduced into a Minimum-Variance Distortionless Response (MVDR) beamformer, {and this paper extends the mechanism to work with a blind source separation algorithm. }
To incorporate dereverberation capability, we further extend {\swIVA} to a blind Convolutional beamforming algorithm ({\swIVAconv}) that integrates {\swIVA} and switching Weighted Prediction Error-based dereverberation ({\swWPE}) in a jointly optimal way. With {\swIVAconv}, two different time-varying characteristics of an observed signal are captured for dereverberation and source separation to achieve effective estimation. We show that both {\swIVA} and {\swIVAconv} can be optimized effectively based on blind signal processing, and their performance can be further improved using a spatial guide for initialization. Experiments demonstrate that {both the} proposed methods largely outperformed conventional {\IVA} and its convolutional beamforming extension ({\IVAconv}) in terms of objective signal quality and automatic speech recognition scores when using  relatively few microphones.
\end{abstract}


\section{Introduction}\label{sec:intro}
When a speech signal is captured by distant microphones, e.g., in a conference room, it often contains such interference signals as reverberation, diffuse noise, and voices from extraneous speakers. They all reduce the intelligibility of the captured speech and often cause serious degradation in many speech applications, such as hands-free teleconferencing and Automatic Speech Recognition (ASR) \cite{robustasr}.

Blind source separation (BSS) has been actively studied to minimize the aforementioned detrimental effects in acquired signals. It separates a sound mixture captured by a microphone array into a given number of sources based on the general statistical characteristics of sources without relying on the prior knowledge of the individual sources or the recording conditions. 
A number of techniques have been developed for BSS, including Independent Component Analysis (ICA) \cite{common,ica}, Independent Vector Analysis (IVA) \cite{iva,hiroeiva,IP}, Full-rank spatial Covariance Analysis (FCA) \cite{Ngoc2010taslp,Ito2021taslp}, and spatial clustering-based beamforming \cite{sawada11aslp,tranvu,souden10aslp}. 
Among them, IVA separates sources as mutually independent vectors, each of which corresponds to each source's full-band complex spectra in the Short-Time Fourier Transform (STFT) domain. 
Due to this mechanism, IVA can solve BSS's common frequency permutation problem \cite{sawada11aslp} without any post processing. In addition, IVA assumes a determined condition for BSS, i.e., the number of sources $N$ equals that of microphones $M$. Under this assumption and when we have more microphones than sources, i.e., $M>N$, it is empirically confirmed that IVA can separate $N$ sources and $M-N$ noise components from a sound mixture with stationary diffuse noise \cite{Sawada2005ICASSP,YukiKubo}. Based on this capability, Independent Vector Extraction (IVE) was recently proposed as a variation of IVA that can extract only $N$ sources from such a noisy mixture in a computationally efficient way \cite{Koldovsky2018taslp,Robin2019waspaa,ikeshita2021taslp}.

For performing blind dereverberation (BD), Weighted Prediction Error-based dereverberation (WPE) has been effective \cite{wpe,gwpe,swpe}. It uses a Multi-Channel Linear Prediction (MCLP) filter to perform dereverberation, and optimizes the filter based on Maximum Likelihood (ML) estimation. WPE can improve such signal processing techniques as beamforming\cite{yoshioka2013eusipco,GSS}, ASR\cite{delcroix15eurasip,SPM}, and speaker recognition/diarization\cite{Voices2019,horiguchi2021arxiv} by using it as preprocessing in far-field signal capturing situations.

Convolutional BeamFormers (CBFs) \cite{Douglas,trinicon,takuya2011taslp,Togami2012icassp,Kagami2018icassp,Gannot2019,ikeshita2019waspaa,Dietzen2020} have also been studied to jointly perform BSS and BD. A CBF is defined in this paper as an STFT-domain filter that spans more than one time frame. It can be factorized into a separation matrix and an MCLP filter and perform BSS and BD in a jointly optimal way by integrating IVA and WPE \cite{takuya2011taslp,Kagami2018icassp,nakatani2020interspeech}. This paper refers to this type of CBF optimization algorithm as a 
\textit{blind Convolutional beamforming algorithm with IVA ({\IVAconv})}. 
Computationally efficient optimization techniques have been developed for {\IVAconv} with factorization\cite{nakatani2020interspeech}, or without it\cite{ikeshita2019waspaa,Nakashima2021icassp}. 
An extension that incorporates IVE into {\IVAconv} has also been proposed \cite{nakatani2021icassp,ikeshita2021spl,togami2020apsipa} that works very well to jointly perform denoising, dereverberation, and source separation when $M\gg N$, e.g., $M=8$ and $N=2$ \cite{nakatani2021icassp}.

\begin{table}[t]
\setlength{\tabcolsep}{5pt}
    \centering
    \caption{Taxonomy of algorithms for dereverberation (DR), beamforming (BF), and convolutional beamforming (CBF)}    \label{tab:taxonomy}
    \begin{tabular}{|c|c|c|c|}\hline
        \rule{0pt}{2.ex}\multirow{2}{*}{\parbox{0.9cm}{\centering DR/BF/\\CBF}} 
        & \multirow{2}{*}{Switch} 
        & \multirow{2}{*}{\parbox{2.5cm}{\centering With ATF}}
        & \multirow{2}{*}{\parbox{2.5cm}{\centering Blind }} \\
        \rule{0pt}{2.ex}  &   &   & \\\hline
        \rule{0pt}{2.3ex}DR  & - & - & WPE \cite{wpe,gwpe,swpe}\\        \rule{0pt}{2.3ex}BF  & - & MVDR BF \cite{Veen88ASSP} & IVA \cite{iva,hiroeiva,IP}\\
        \rule{0pt}{2.3ex}CBF & - & wMPDR CBF \cite{factorizedwpd,nakatani2020taslp} & {\IVAconv} \cite{ikeshita2019waspaa,nakatani2020interspeech}\\\hline
        \rule{0pt}{2.3ex}DR  & \checkmark & - & swWPE \cite{SWWPE2021}\\
        \rule{0pt}{2.3ex}BF  & \checkmark & MVDR swBF \cite{yamaoka2019icassp}& swIVA (proposed)\\
        \rule{0pt}{2.3ex}CBF & \checkmark & wMPDR swCBF \cite{EUSIPCO2021} & {\swIVAconv} (proposed)\\\hline
    \end{tabular}
\end{table}

\subsection{Problem solved in this paper}
Despite the success of {\IVA}/{\IVAconv} approaches, achieving high estimation accuracy is still a challenging problem when using just a few microphones in the presence of diffuse noise. The performance largely degrades as the number of microphones $M$ (or more specifically $M-N$) decreases. This severely limits the application area of {\IVA}/{\IVAconv} because using many microphones is usually unacceptable in practical applications.

Recently, a promising algorithm called Minimum Variance Distortionless Response (MVDR) switching BeamFormer (swBF) \cite{yamaoka2019icassp} has been proposed that achieves higher estimation accuracy with a small number of microphones. This algorithm is an extension of a conventional MVDR BeamFormer (BF) \cite{Veen88ASSP,MPDR}. 
MVDR swBF is composed of a set of time-invariant BFs and a switch, which selects one of the BF outputs at each time frame that most accurately estimates the target signal. It relies on the sparseness property of sources, i.e., sources are sparsely distributed in the STFT domain. With this property, the number of sources at each time-frequency (TF) point can be smaller than their total number over all TF points, and thus we can improve beamforming by appropriately switching the BFs. The BFs and the switch are jointly optimized by minimizing the noise power in the observed signal under a distortionless constraint with the given acoustic transfer functions (ATFs) from sources to microphones.  WPE with a switching mechanism called {\swWPE} has also been confirmed effective \cite{SWWPE2021}. It consistently outperformed a conventional WPE in diffuse noise environments and/or with underdetermined conditions. A Weighted Minimum Power Distortionless Response (wMPDR) CBF with a switching mechanism \cite{EUSIPCO2021}, called wMPDR swCBF, also outperformed the conventional wMPDR CBF \cite{nakatani2020taslp}.

\subsection{{Contributions}}

In this paper, we propose two new methods, \textit{switching {\IVA} ({\swIVA})} and \textit{switching {\IVAconv} ({\swIVAconv})}, by respectively incorporating a switching mechanism into {\IVA} and {\IVAconv},\footnote{Although it is straightforward to further incorporate IVE into {\swIVA}/{\swIVAconv}, this paper omits this explanation for conciseness.} to improve their performance when using a small number of microphones. 
We show that the optimization algorithm can be derived under the same assumptions as those for {\IVA} and {\IVAconv}, and all the filter coefficients and switches are jointly optimized based on the maximum-likelihood (ML) estimation. 
{In BSS, certain related techniques have been proposed, e.g., for modeling a time-varying mixing system with a hidden Markov model \cite{glimpsingiva} or handling abrupt changes in sensor or source positions using shadow separation filters \cite{shadowbss}. In contrast, our switching mechanism is incorporated into a separation system of BSS, and the whole system is optimized based on the same objective as the conventional BSS in a unified manner.}
Table~\ref{tab:taxonomy} summarizes the taxonomy of the conventional and proposed algorithms for dereverberation, beamforming, and convolutional beamforming {with and without the switching mechanism}.

\begin{figure}
    \centering
        \includegraphics[width=7.4cm]{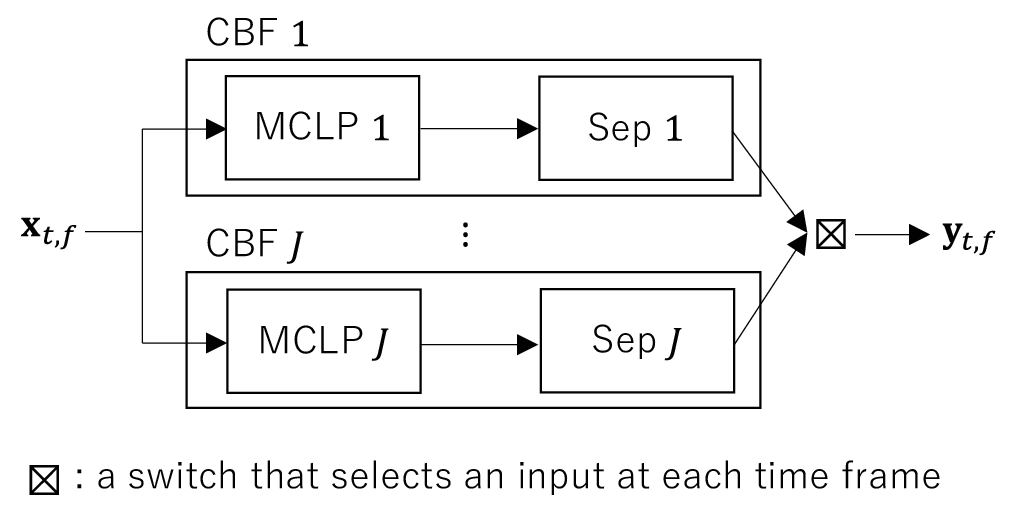}\smallskip\\
        \parbox{7.8cm}{\footnotesize (a) Direct switching model of a swCBF,  composed of a set of CBFs followed by a switch. {Each CBF can be further factorized into an MCLP filter and a separation matrix.}}\bigskip\\ 
        \includegraphics[width=8.5cm]{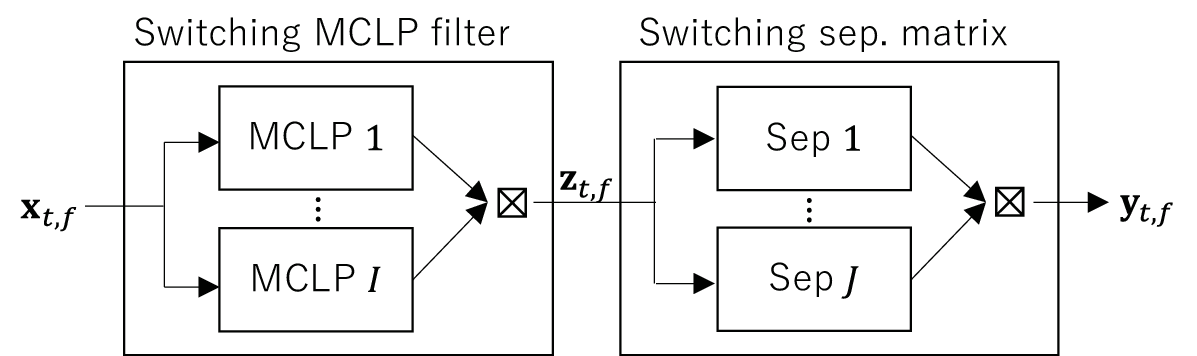}\bigskip\\
        \parbox{8cm}{\footnotesize (b) Factorized switching model of a swCBF,  composed of a switching MCLP filter and a switching separation matrix. }\bigskip
    \caption{Examples of swCBF optimized by {\swIVAconv}: {$\vect{x}_{t,f}$, $\vect{z}_{t,f}$, and $\vect{y}_{t,f}\in\mathbb{C}^M$ represent vectors of a multichannel observed signal, a multichannel dereverberated signal, and swCBF output at a TF point $(t,f)$.}}
    \label{fig:my_label2}
\end{figure}

{This paper further proposes several important techniques to make {\swIVA} and {\swIVAconv} work effectively. The first one is the structure of a swCBF used by {\swIVAconv}. One may adopt a simple structure, called \textit{a direct switching model} (Fig.~1(a)), composed of a set of CBFs followed by a single switch that selects one of the CBF outputs at each time frame. However, as shown in our experiments (Section~\ref{exp:switches}), this model is sub-optimal.} Instead, we propose \textit{a factorized switching model} (Fig.~1(b)) that uses two different switches for MCLP filters and separation matrices. This model is advantageous for separately capturing two different time-varying characteristics of an observed signal for dereverberation and source separation.  

{The second important technique is the structure of a swBF.} With a conventional swBF \cite{EUSIPCO2021}, individual sources are estimated separately by different swBFs, each of which independently selects one of its BFs for the estimation at each time frame.  
{With this structure, however, it is difficult to blindly optimize {\swIVA} in a computationally efficient way.} To overcome this problem, we propose a new switching structure, called \textit{separation matrix-wise switching}, which selects one of the separation matrices, each of which estimates all the sources at once at each time frame (right half of Fig.~1(a)). It allows us to also use computationally efficient optimization techniques proposed for conventional {\IVA} \cite{IP,ISS,IPA,Andreas2021icassp} for {\swIVA}.

{The third technique is about} the selection of a statistical source model. Although a frequency-dependent source model is advantageous for WPE and the optimization of switches, a frequency-independent source model is essential for {\IVA}. As a practical technique to solve this discrepancy, we introduce a hybrid source model, referred to as a \textit{coarse-fine source model}. With it, we use a frequency-independent model to update the separation matrix and a frequency-dependent model to update the other parameters, including the switch.

{The final technique concerns initialization.} As shown in our experiments, simple initialization, which is often used for conventional {\IVA}, does not work well for {\swIVA}. This is partly because sources are estimated using different separation matrices at individual switching states with {\swIVA}, and thus they may be permuted at different states. We call this an inter-state permutation problem. To solve it, we propose two effective initialization techniques: 1) \textit{blind single-state initialization} and 2) \textit{spatially guided initialization}. The former allows us to optimize {\swIVA} and {\swIVAconv} by complete blind processing. The latter uses a TF mask-based BF estimated, e.g., by a neural network (NN) \cite{nakatani2020taslp} to initialize the separation matrices. The spatial guide not only solves the inter-state permutation but also improves the optimization to converge to better stationary points \cite{boedekker2021arxiv}.

In experiments, we evaluated {\swIVA}/{\swIVAconv} in terms of signal distortion reduction and ASR improvement using noisy reverberant speech mixtures. We first show that the above proposed techniques are all essential for {\swIVA} and {\swIVAconv} to perform effective estimation. With these techniques, the switching mechanism largely improves the performances of {\swIVA} and {\swIVAconv}.  The improvement is substantial and consistent under all experimental conditions when the number of switching states is increased from one (i.e., {\IVA} or {\IVAconv}) to two (i.e., {\swIVA} or {\swIVAconv}) for both the separation matrices and the MCLP filters, although it becomes less stable when additionally increasing the number of states.

In the remainder of this paper, Section~\ref{sec:problem} defines the problem and Section~\ref{sec:baseline} presents a formulation of {\swIVA}/{\swIVAconv}.  
In Sections~\ref{sec:spguide} and V, we describe effective initialization techniques and the computational time complexity of the proposed methods. Experiments and concluding remarks are given in Sections \ref{sec:experiments} and \ref{sec:conclude}.

\section{{Problem in conventional methods}}\label{sec:problem}
{This section elaborates the problem mentioned in Section I.A. After describing the model of the observed signal, we discuss the problem to be solved in the conventional methods. }

{In this section and the next, we mainly discuss cases with reverberation and develop a method based on {\IVAconv}. Then in Section \ref{sec:swIVA}, we limit the discussion to cases without reverberation and present a method based on {\IVA}.}

\subsection{Model of observed signals}\label{sec:obsmodel}
Suppose that $N$ speech signals are captured by $M$ distant microphones with reverberation and background diffuse noise. We assume $M\ge N$ in this paper. 
Let $x_{m,t,f}$ be the captured signal at the $m$th microphone and a TF point $(t,f)$ in the STFT domain for $1\le t\le T$ and $1\le f\le F$, where $T$ and $F$ are the numbers of time frames and frequency bins, and let $(\cdot)^{\top}$ denote a non-conjugate transpose. Then the captured signal at all the microphones, $\vect{x}_{t,f}=[x_{1,t,f},\ldots,x_{M,t,f}]^{\top}
\in\mathbb{C}^{M}$, is modeled by
\begin{align}
    \vect{x}_{t,f}&=\sum_{n=1}^N\vect{d}_{n,t,f}+\sum_{n=1}^N\vect{l}_{n,t,f}+\vect{v}_{t,f},\label{eq:obs1}\\
    \vect{d}_{n,t,f}&=\vect{h}_{n,f}s_{n,t,f}
    \quad\mbox{for all}~n,\label{eq:obs2}
\end{align}
where $\vect{d}_{n,t,f}=[d_{n,1,t,f},\ldots,d_{n,M,t,f}]^{\top}\in\mathbb{C}^M$ is the direct signal plus the early reflections of the $n$th source \cite{early,Nishiura2007interspeech}, $\vect{l}_{n,t,f}$ is the source's late reverberation, and $\vect{v}_{t,f}$ is the diffuse noise. This paper deals with $\vect{d}_{n,t,f}$ for each $n$ as a signal to be estimated, called a desired signal, and models it by a product of a time-invariant ATF $\vect{h}_{n,f}\in\mathbb{C}^{M}$ and the $n$th clean source signal $s_{n,t,f}\in\mathbb{C}$ in Eq.~(\ref{eq:obs2}). 

\subsection{Problem in conventional {\IVAconv}}\label{sec:conv}

To perform BSS and BD, the conventional {\IVAconv} assumes determined conditions,\footnote{To be strict, an MCLP filter requires overdetermined conditions, where microphones outnumber point sources, for performing precise dereverberation according to the multiple-input/output inverse theorem (MINT) \cite{mint}.} where the captured signal contains only point sources, and the number of sources equals the number of microphones. This situation causes substantial mismatch with the above observed signal model. Since diffuse noise comes to microphones from arbitrary directions, it is more appropriate to model the signal to be composed of many sources. Accordingly, to formulate {\IVAconv}, we need to further simplify the observation model, that is, modeling it by composing it of $N$ speech sources and $M-N$ noise components \cite{ikeshita2021taslp}.
In practice, this simplification works well when $M\gg N$ (e.g., $M=8$ and $N=2$) and the noise level is moderate.

With the above simplification, {\IVAconv} applies a CBF to the observed signal:
\begin{align}
    \vect{y}_{t,f}&
    =\left[\begin{array}{c}
    \vect{W}_f\\\bar{\vect{W}}_f
    \end{array}\right]^{\HT}
    \left[\begin{array}{c}
    \vect{x}_{t,f}\\\bar{\vect{x}}_{t,f}
    \end{array}\right]
    \in\mathbb{C}^{M},\label{eq:tiCBF}\\
    \bar{\vect{x}}_{t,f}&=[\vect{x}_{t-D,f}^{\top},\ldots,\vect{x}_{t-L+1,f}^{\top}]^{\top}\in\mathbb{C}^{M(L-D)},\nonumber
\end{align}
where $(\cdot)^{\HT}$ is an Hermitian transpose,  $\vect{W}_f\in\mathbb{C}^{M\times M}$ and $\bar{\vect{W}}_f\in\mathbb{C}^{M(L-D)\times M}$ are the CBF's coefficient matrices applied to current captured signal $\vect{x}_{t,f}$ and past captured signal sequence $\bar{\vect{x}}_{t,f}$, {and $L$ is the length of a CBF}.
With a CBF, $\bar{\vect{W}}_f$ is introduced to appropriately handle reverberation that is longer than an analysis window.
$\vect{y}_{t,f}=[y_{1,t,f},\ldots,y_{M,t,f}]^{\top}\in\mathbb{C}^{M}$ is the CBF output, including $N$ signals that correspond to estimates of the desired signals.
The other $M$ signals in $\vect{y}_{t,f}$ correspond to estimated noise components.  $D~(\ge 1)$ is {a} prediction delay introduced to set the dereverberation goal to reduce only the late reverberation and preserve the desired signals \cite{wpe}. 

Although the observed signal can be well dereverberated and separated into $N$ speech signals and $M-N$ noise components when $M\gg N$, the estimation becomes challenging as $M-N$ becomes small and/or the noise level becomes large. A relatively high level of noise and large estimation errors remain in the separated speech signals, and the dereverberation capability is seriously degraded. 
This problem severely limits the applicability of {\IVAconv} especially when the number of available microphones is small (e.g., 2 or 3).

\section{Formulation of {\swIVA}/{\swIVAconv}}\label{sec:baseline}
In this section, we first describe our motivation for introducing a switching mechanism to solve the problem in the conventional methods and then present the formulation of {\swIVA}/{\swIVAconv} by introducing a switching mechanism into IVA/{\IVAconv}.

\subsection{Motivation for introducing a switching mechanism}\label{sec:motivation}

The main aim of introducing a switching mechanism to {\IVAconv} is to reduce the model mismatch discussed in Section~\ref{sec:problem}-B and to improve the estimation accuracy, especially when using a relatively small number of microphones. 
Our expectation is that the mismatch can be reduced by clustering the time frames of an observed signal into several groups so that each group better fits the assumed observation model with given set of microphones.
Then, each group with reduced mismatch could be more accurately handled by a conventional CBF. 
In general, the signal components included in an observed signal, such as speech signals and their reverberation, are highly time-varying, and active components rapidly change in a frame-by-frame manner. Therefore, adaptively changing the filter coefficients so that they better fit the active components at each TF point can be advantageous over conventional approaches that use {fixed filter coefficients} to handle all the signal components over time at once. The switching mechanism performs such adaptive processing by clustering the time frames of an observed signal into groups with reduced mismatch and individually applying a conventional CBF to each of the groups.

Let us present two examples that explain how the switching mechanism reduces mismatch. 
First, a speech signal has a sparseness property, i.e., the energy of a speech signal is sparsely distributed in the STFT domain, and different signals rarely overlap on each TF point. According to this property, clustering can be performed to reduce the number of sources $N$ included in each group \cite{yamaoka2019icassp}, and thus increased signal space $M-N$ can be used to improve the separation of diffuse noise in each group. 
Next, for dereverberation, as will be described in the following subsections, prediction matrices are used to estimate the late reverberation included in each time frame from a past observed signal sequence. In a diffuse noise environment, the prediction matrices also need to reduce the influence of the noise in the past observed signal sequence on the prediction \cite{wpe}. Because such influence can be highly time-varying, the prediction matrices that optimally reduce the influence should also be time-varying. With the switching mechanism, such time-varying prediction matrices could be achieved 
by clustering the observed signals considering the influence of the noise, and applying different sets of prediction matrices to the clustered groups. 

\subsection{Definition of swCBF}\label{sec:formulation}
Before deriving {\swIVAconv}, this subsection first defines {a} structure of a CBF with a switching mechanism: a swCBF. 

We start by defining a time-varying CBF, which we modify into a swCBF. A time-varying CBF is defined simply by letting the coefficient matrices of a CBF in Eq.~(\ref{eq:tiCBF}) be time-varying:
\begin{align}
    \vect{y}_{t,f}&
    =\left[\begin{array}{c}
    \vect{W}_{t,f}\\\bar{\vect{W}}_{t,f}
    \end{array}\right]^{\HT}
    \left[\begin{array}{c}
    \vect{x}_{t,f}\\\bar{\vect{x}}_{t,f}
    \end{array}\right]
    \in\mathbb{C}^{M},\label{eq:tvCBF}
\end{align}
where $\vect{W}_{t,f}$ and $\bar{\vect{W}}_{t,f}$ are time-varying CBF coefficients. 
Then, similar to a conventional CBF \cite{takuya2011taslp,nakatani2020interspeech,ikeshita2021spl}, we define a  factorized form of the time-varying CBF:
\begin{align}
    \left[\begin{array}{c}
    \vect{W}_{t,f}\\\bar{\vect{W}}_{t,f}
    \end{array}\right] = \left[
    \begin{array}{c}\vect{I}_M\\-\vect{G}_{t,f}\end{array}
    \right]\vect{W}_{t,f},\label{eq:fact}
\end{align}
where $\vect{G}_{t,f}\in\mathbb{C}^{M(L-D)\times M}$ is a coefficient matrix that satisfies $\bar{\vect{W}}_{t,f}=-\vect{G}_{t,f}\vect{W}_{t,f}$, and $\vect{I}_M\in\mathbb{R}^{M\times M}$ is an identity matrix. Using this factorization, $\vect{y}_{t,f}$ in Eq.~(\ref{eq:tvCBF}) is obtained:
\begin{align}
    \vect{z}_{t,f}&=\vect{x}_{t,f}-\vect{G}_{t,f}^{\HT}\bar{\vect{x}}_{t,f},\label{eq:wpe}\\
    \vect{y}_{t,f}&=\vect{W}_{t,f}^{\HT}\vect{z}_{t,f},\label{eq:sep}
\end{align}
where Eq.~(\ref{eq:wpe}) is an MCLP filter that yields dereverberated signal $\vect{z}_{t,f}$ from $\vect{x}_{t,f}$ using prediction matrix $\vect{G}_{t,f}$ and Eq.~(\ref{eq:sep}) extracts $\vect{y}_{t,f}$ by applying separation matrix $\vect{W}_{t,f}$ to $\vect{z}_{t,f}$.  Eqs.~(\ref{eq:wpe}) and (\ref{eq:sep}) correspond to filters used by {\swWPE} and {\swIVA}. 

Because the above time-varying CBF is so flexible that over-fitting to the observed signal can easily happen, we need to introduce certain constraints to avoid that situation. For this purpose, we introduce a switching mechanism, called a factorized switching model in Fig.~1(b), to the MCLP filter and the separation matrix. It is composed of a set of time-invariant MCLP filters and separation matrices, which are controlled by their respective switches. We mathematically model them by the sums of the time-invariant coefficient matrices with switching weights:
\begin{eqnarray}
    \hspace{-6mm}&\displaystyle\vect{G}_{t,f}=\sum_{i=1}^{I}\gamma^{(i)}_{t,f}\vect{G}^{(i)}_{f}~~\mbox{and} ~~\vect{W}_{t,f}=\sum_{j=1}^J\delta^{(j)}_{t,f}\vect{W}^{(j)}_{f},\label{eq:switching}\\
    \hspace{-6mm}&\displaystyle\gamma^{(i)}_{t,f}\in\{0,1\},~\delta^{(j)}_{t,f}\in\{0,1\},~\sum_{i=1}^I\gamma^{(i)}_{t,f}=1,~\sum_{j=1}^J\delta^{(j)}_{t,f}=1, 
\end{eqnarray}
where $I$ and $J$ are the numbers of the switching states of the MCLP filter and the separation matrix, $\vect{G}^{(i)}_{f}$ for $1\le i\le I$ is a prediction matrix of the $i$th time-invariant MCLP filter,  $\vect{W}^{(j)}_{f}$ for $1\le j\le J$ is the $j$th time-invariant separation matrix, and  
$\{\gamma^{(i)}_{t,f}\}_{i,t,f}$ and $\{\delta^{(j)}_{t,f}\}_{j,t,f}$ are their time-varying switching weights. In this paper, for brevity, we only consider hard switches and allow $\gamma^{(i)}_{t,f}$ and $\delta^{(j)}_{t,f}$ to take only binary values, 0 or 1.

\begin{figure}
    \centering
    \includegraphics[width=8.cm]{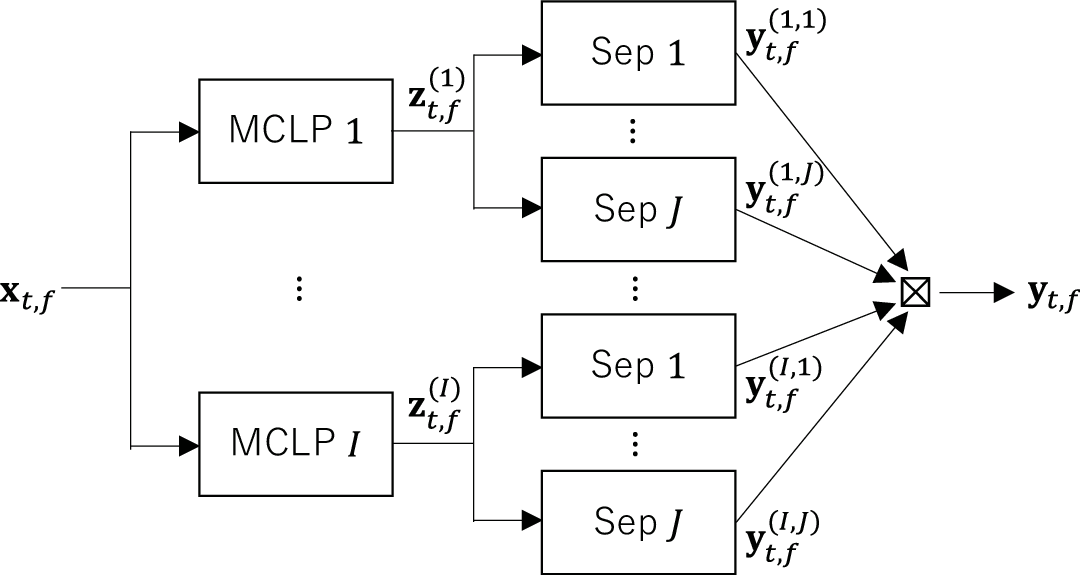}
    \caption{Expanded form of a swCBF: Separation matrices (Sep) labeled with identical number share same filter coefficients.}
    \label{fig:expform}
\end{figure}

Based on Eqs.~(\ref{eq:wpe}), (\ref{eq:sep}), and (\ref{eq:switching}), the swCBF output $\vect{y}_{t,f}$ can be further rewritten:
\begin{align}
    \vect{z}^{(i)}_{t,f}&=\vect{x}_{t,f}-(\vect{G}^{(i)}_{f})^{\HT}\bar{\vect{x}}_{t,f},\label{eq:zupdate}\\
    \vect{y}^{(i,j)}_{t,f}&=(\vect{W}^{(j)}_{f})^{\HT}\vect{z}^{(i)}_{t,f}\label{eq:yupdate},\\
    \vect{y}_{t,f}&=\sum_{i=1}^I\sum_{j=1}^J\beta^{(i,j)}_{t,f}\vect{y}^{(i,j)}_{t,f}\label{eq:ysum},\\
    \beta^{(i,j)}&\in\{0,1\},~\sum_{i=1}^I\sum_{j=1}^J\beta^{(i,j)}_{t,f}=1,
\end{align}
where $\beta^{(i,j)}_{t,f}=\gamma^{(i)}_{t,f}\delta^{(j)}_{t,f}$ is a unified switching weight that takes a binary value, 0 or 1. We call this an expanded form of a swCBF. 
Figure~\ref{fig:expform} also illustrates the structure of the form. 
In the following, we use this form to derive the optimization algorithm.

{With the expanded form in Fig.~\ref{fig:expform}, at each TF point, multichannel observed signal $\vect{x}_{t,f}$ is first dereverberated by $I$ different MCLP filters using Eq.~(\ref{eq:zupdate}) to yield $I$ different sets of dereverberated signals $\vect{z}_{t,f}^{(i)}$, which are separated by $J$ different separation matrices using Eq.~(\ref{eq:yupdate}) to yield $IJ$ different output signals $\vect{y}_{t,f}^{(i,j)}$. Finally, a switch selects one of the $IJ$ output signals using Eq.~(\ref{eq:ysum}). To perform accurate time-varying filtering, all the filter coefficients and switching weights are optimized based on the ML objective derived in the following subsections.}

One may consider another naive structure of a swCBF, called a direct switching model (Fig.~1(a)).\footnote{Because a swCBF with the direct switching model can also be represented by the expanded form, the optimization algorithm developed in this paper can also be applied to it.} It is composed of $J$ CBFs and a switch that selects one of the CBF's outputs at each time frame, and each CBF may be further decomposed into a pair of an MCLP filter and a separation matrix. This model, however, is sub-optimal for {\swIVAconv} as will be shown in experiments because it cannot separately capture the time-varying characteristics of the signals for dereverberation and source separation.

\subsection{Probabilistic model for {\swIVAconv}}
Now, we introduce the probabilistic models to derive the ML objective. They are equivalent to those used for {\IVAconv}. We assume that a certain desired swCBF satisfies the following conditions:
\begin{enumerate}
    \item {Each element $y_{n,t,f}$ of swCBF output $\vect{y}_{t,f}$} for $1\le n\le M$, $1\le t \le T$ and $1\le f \le F$, is mutually independent, satisfying
    \begin{align}
        p(\{y_{n,t,f}\}_{n,t,f})&=\prod_{n=1}^M\prod_{t=1}^{T}\prod_{f=1}^Fp(y_{n,t,f}),\label{eq:independence}
    \end{align}
    {where $n$ is an index of separated sources $y_{n,t,f}$.}
    \item Each $y_{n,t,f}$ for $1\le n\le M$ can be modeled by a time-varying Gaussian with a mean zero and time-varying variance $\lambda_{n,t,f}$:
\begin{align}
    p(y_{n,t,f};\lambda_{n,t,f})=\frac{1}{\pi\lambda_{n,t,f}}\exp\left(-\frac{|y_{n,t,f}|^2}{\lambda_{n,t,f}}\right).\label{eq:coarse}
\end{align}
\end{enumerate}
{Note here that we assumed the mutual independence of $y_{n,t,f}$ between different frequencies in Eq.~(\ref{eq:independence}), and adopted the frequency-dependent source model in Eq.~(\ref{eq:coarse}) for swCIVA formulation.  With this setting, however, BSS solutions are known to suffer from the frequency permutation problem.  As a practical technique to solve it, this paper below introduces a frequency-independent source model, which is often adopted by the conventional {\IVA}, in Section~\ref{sec:cfmodel}.}

Based on the above two models {and according to Appendix~A}, we obtain the log likelihood function for a time-varying CBF, 
disregarding constant terms:
\begin{align}
    {\cal L}\left(\theta\right)=&-\sum_{n=1}^M\sum_{t=1}^{T}\sum_{f=1}^F\left(\frac{|y_{n,t,f}|^2}{\lambda_{n,t,f}}+\log\lambda_{n,t,f}\right)\nonumber\\
    & +2\sum_{t=1}^{T}\sum_{f=1}^F\log|\det\vect{W}_{t,f}|,
\end{align}
where $\theta=\{\{\lambda_{n,t,f}\}_{n,t,f},\{\vect{W}_{t,f}\}_{t,f},\{\bar{\vect{W}}_{t,f}\}_{t,f}\}$, and $y_{n,t,f}$ is obtained by Eq.~(\ref{eq:tvCBF}) dependent on $\vect{W}_{t,f}$ and $\bar{\vect{W}}_{t,f}$. 

Finally, with the expanded form defined by Eqs.~(\ref{eq:zupdate}) to (\ref{eq:ysum}), the likelihood function of {\swIVAconv} can be written:
\begin{align}
    &\hspace{-.5mm}{\cal L}\left({\cal G}, {\cal W}, \Lambda, {\cal B}\right)=\sum_{t,f,i,j}\beta^{(i,j)}_{t,f}{\cal L}^{(i,j)}_{t,f}\left(\vect{G}^{(i)}_{f},\vect{W}^{(j)}_{f},\Lambda_{t,f}\right),\label{eq:swML1}\\
    &{\cal L}^{(i,j)}_{t,f}\left(\vect{G}^{(i)}_{f},\vect{W}^{(j)}_{f},\Lambda_{t,f}\right)=-\sum_{n=1}^M\left(\frac{|y^{(i,j)}_{n,t,f}|^2}{\lambda_{n,t,f}}+\log\lambda_{n,t,f}\right)\nonumber\\
    &\hspace{4.1cm}+2\log|\det\vect{W}^{(j)}_{f}|,
    \label{eq:swML2}
\end{align}
where $y^{(i,j)}_{n,t,f}$ is 
obtained by Eqs.~(\ref{eq:zupdate}) and (\ref{eq:yupdate}), dependent on $\vect{G}^{(i)}_f$ and $\vect{W}^{(j)}_f$, ${\cal G}=\{\vect{G}^{(i)}_{f}\}_{i,f}$, ${\cal W}=\{\vect{W}^{(j)}_{f}\}_{j,f}$, $\Lambda=\{\Lambda_{t,f}\}_{t,f}$, $\Lambda_{t,f}=\{\lambda_{n,t,f}\}_n$, and ${\cal B}=\{\beta^{(i,j)}_{t,f}\}_{i,j,t,f}$.

\subsection{Optimization algorithm {\swIVAconv}}\label{sec:optimization}
We now derive the algorithm, {\swIVAconv}, which optimizes a swCBF defined by Eqs.~(\ref{eq:zupdate}), (\ref{eq:yupdate}), and (\ref{eq:ysum}) based on the ML objective defined by Eqs.~(\ref{eq:swML1}) and (\ref{eq:swML2}). 

Because no closed form solution has been obtained for the optimization, we use iterative estimation based on a coordinate ascent method\cite{CD2015}. It alternately updates each parameter subset by fixing the other parameter subsets and iterates the update until a convergence is obtained. 
The following describes each update step in the iteration.

\subsubsection{${\cal G}$ update}

First, we extract the terms related with $\vect{G}^{(i)}_{f}$ from Eqs.~(\ref{eq:swML1}) and (\ref{eq:swML2}) and obtain
\begin{align}
    {\cal L}_{\vect{G}^{(i)}_{f}}
    &=-\sum_{n,j,t}\frac{\beta^{(i,j)}_{t,f}}{\lambda_{n,t,f}}
    \left|\left(\vect{w}^{(j)}_{n,f}\right)^{\HT}\left(\vect{x}_{t,f}-(\vect{G}^{(i)}_{f})^{\HT}\bar{\vect{x}}_{t,f}\right)\right|^2,
\end{align}
where $\vect{w}^{(j)}_{n,f}$ is the $n$th column of $\vect{W}^{(j)}_{f}$. Since the above equation is a simple quadratic form in terms of $\vect{G}^{(i)}_{f}$, we can obtain a closed form solution for it when fixing the other parameters.  
Let $\vect{g}^{(i)}_{f}=\mbox{vec}(\vect{G}^{(i)}_{f})$, where $\vect{a}=\mbox{vec}(\vect{A})$ is an operation to reshape matrix $\vect{A}=[\vect{a}_1,\ldots,\vect{a}_M]$ to vector $\vect{a}=[\vect{a}_1^{\top},\ldots,\vect{a}_M^{\top}]^{\top}$.
Then the solution is given by
\begin{align}
    \vect{g}^{(i)}_{f}&\leftarrow(\Psi^{(i)}_{f})^{-1}\mbox{vec}(\Phi^{(i)}_{f})\in\mathbb{C}^{M^2(L-D)}.\label{eq:gupdate}
\end{align}
where $(\cdot)^{-1}$ denotes a matrix inversion. $\Psi^{(i)}_{f}\in\mathbb{C}^{M^2(L-D)\times M^2(L-D)}$ and $\Phi^{(i)}_{f}\in\mathbb{C}^{M(L-D)\times M}$ are calculated:
\begin{align}
    \Psi^{(i)}_{f}&=\sum_{j=1}^J\sum_{n=1}^M\left(\vect{w}^{(j)}_{n,f}(\vect{w}^{(j)}_{n,f})^{\HT}\right)^*\otimes\vect{R}^{(i,j)}_{n,f},\\
    \Phi^{(i)}_{f}&=\sum_{j=1}^J\sum_{n=1}^M\vect{P}^{(i,j)}_{n,f}\left(\vect{w}^{(j)}_{n,f}(\vect{w}^{(j)}_{n,f})^{\HT}\right),
\end{align}
where $(\cdot)^*$ is a complex conjugate, $\otimes$ is a Kronecker product, and $\vect{R}^{(i,j)}_{n,f}\in\mathbb{C}^{M(L-D)\times M(L-D)}$ and $\vect{P}^{(i,j)}_{n,f}\in\mathbb{C}^{M(L-D)\times M}$ are spatio-temporal covariance matrices for the $n$th source, obtained by
\begin{align}
    \vect{R}^{(i,j)}_{n,f}&=\sum_{t=1}^{T}\frac{\beta^{(i,j)}_{t,f}}{\lambda_{n,t,f}}\bar{\vect{x}}_{t,f}\bar{\vect{x}}_{t,f}^{\HT},\label{eq:Rupdate}\\
    \vect{P}^{(i,j)}_{n,f}&=\sum_{t=1}^{T}\frac{\beta^{(i,j)}_{t,f}}{\lambda_{n,t,f}}\bar{\vect{x}}_{t,f}\vect{x}_{t,f}^{\HT}.\label{eq:Pupdate}
\end{align}
Note that the above update steps can be viewed as an extension of {\swWPE} \cite{SWWPE2021} using a spatial model specified by ${\cal W}$. 

{Once $\vect{G}_{f}^{(i)}$ is updated, $\{\vect{z}_{t,f}^{(i)}\}_{i,t,f}$ is also updated by Eq.~(\ref{eq:zupdate}).}

{Several computationally efficient algorithms have been proposed for the joint optimization of {\IVAconv} \cite{ikeshita2019waspaa,nakatani2020interspeech,Nakashima2021icassp,nakatani2021icassp,ikeshita2021spl}. Among them, the above update steps correspond to a technique called source-wise covariance decomposition \cite{nakatani2020interspeech}, adopted by this paper because the other algorithms cannot be used with the factorized switching model. Although the adopted algorithm requires the calculation of relatively large covariance matrices, $\Psi_f^{(i)}$ and $\Phi_f^{(i)}$, the increase of the computational cost owing to it is not necessarily significant in practice when the number of microphones is relatively small \cite{nakatani2020interspeech}.  In contrast, when we adopt the direct switching model, we can use all the computationally efficient algorithms. However, the model underperforms the factorized switching model in terms of estimation accuracy as will be shown in our experiments.}

\subsubsection{${\cal W}$ update}
Extracting terms related with $\vect{W}^{(j)}_{f}$ from Eqs.~(\ref{eq:swML1}) and (\ref{eq:swML2}) yields
\begin{align}
    {\cal L}_{\vect{W}^{(j)}_{f}}
    &=-\sum_{n=1}^M(\vect{w}^{(j)}_{n,f})^{\HT}{\Sigma^{(j)}_{n,f}}\vect{w}^{(j)}_{n,f}+2T^{(j)}_{f}\log|\det\vect{W}^{(j)}_{f}|,\label{eq:wobj}\\
    \Sigma^{(j)}_{n,f}&=\sum_{i=1}^I\sum_{t=1}^{T}\frac{\beta^{(i,j)}_{t,f}}{\lambda_{n,t,f}}\vect{z}^{(i)}_{t,f}(\vect{z}^{(i)}_{t,f})^{\HT},\label{eq:qcov}\\
    T^{(j)}_{f}&=\sum_{i=1}^I\sum_{t=1}^{T}\beta^{(i,j)}_{t,f},
\end{align}
where $\vect{z}^{(i)}_{t,f}$ is the output of the $i$th MCLP filter in Eq.~(\ref{eq:zupdate}). Because the above objective has the same form as {\IVA} \cite{IP}, we can apply iterative optimization techniques proposed for it, including Iterative Projection (IP) \cite{IP}, Iterative Source Steering \cite{ISS}, Iterative Projection with Adjustment (IPA) \cite{IPA}, and accelerated AuxIVA \cite{Andreas2021icassp}. This paper employs IP and updates each beamformer for separating the $n$th source for $1\le n\le M$ (or the $n$th column of $\vect{W}^{(j)}_{f}$):
\begin{align}
    \vect{w}^{(j)}_{n,f}&\leftarrow \left((\vect{W}^{(j)}_{f})^{\HT}\Sigma^{(j)}_{n,f}\right)^{-1}\vect{e}_n,\label{eq:wupdate1}\\
    \vect{w}^{(j)}_{n,f}&\leftarrow\vect{w}^{(j)}_{n,f}/\left((\vect{w}^{(j)}_{n,f})^{\HT}\Sigma^{(j)}_{n,f}\vect{w}^{(j)}_{n,f}\right)^{1/2},\label{eq:wupdate2}
\end{align}
where $\vect{e}_n$ is the $n$th column of identity matrix $\vect{I}_M$. 

{Once $\vect{W}_{f}^{(j)}$ is updated, $\{\vect{y}_{t,f}^{(i,j)}\}_{i,j,t,f}$ is also updated by Eq.~(\ref{eq:yupdate}).}

\subsubsection{$\Lambda$ and ${\cal B}$ updates}
After updating $\vect{y}_{t,f}$ by Eqs.~(\ref{eq:yupdate}) and (\ref{eq:ysum}), variance $\lambda_{n,t,f}$ is updated based on the likelihood function in Eq.~(\ref{eq:swML1}) by the power of $y_{n,t,f}$:
\begin{align}
    \lambda_{n,t,f}\leftarrow|y_{n,t,f}|^2+\varepsilon,\label{eq:lambdaupdate}
\end{align}
where $\varepsilon$ is a small positive scalar to avoid zero division during the optimization.

Then, according to Eq.~(\ref{eq:swML1}), the switching weight can be updated:
\begin{align}
    &\beta^{(i,j)}_{t,f}\leftarrow
    \left\{\begin{array}{ll}
        1 & \hspace{-1mm}\mbox{if~} \{i,j\}=\argmax_{\{i',j'\}} 
        {\cal L}_{t,f}^{(i',j')}\left(\vect{G}^{(i')}_{f},\vect{W}^{(j')}_{f},\Lambda_{t,f}\right)\\
        0 & \hspace{-1mm}\mbox{otherwise},
    \end{array}\right.\label{eq:betaupdate}
\end{align}
{where ${\cal L}_{t,f}(\vect{G}^{(i)}_{f},\vect{W}^{(j)}_{f},\Lambda_{t,f})$ is calculated by Eq.~(\ref{eq:swML2}). }

\subsection{Introduction of coarse-fine source model}\label{sec:cfmodel}

A coarse-fine source model is an important heuristic to ensure that {\swIVAconv} works effectively. 
As will be shown in our experiments, the frequency-dependent source model (a fine source model) defined in Eq.~(\ref{eq:coarse}) is essential for appropriately optimizing the MCLP filters and the switching weights. However, using a frequency-independent source model (a coarse source model) is essential so that IVA can effectively solve the frequency permutation problem. With the coarse-fine source model, we apply the coarse source model to update the separation matrices while using the fine source model for the other part of the optimization.  
The coarse source model is defined at each TF point:
\begin{align}
    p(y_{n,t,f};\lambda_{n,t})=(\pi\lambda_{n,t})^{-1}\exp\left(-\frac{|y_{n,t,f}|^2}{\lambda_{n,t}}\right),
\end{align}
where $\lambda_{n,t}$ is a time dependent and frequency independent variance of the $n$th source.
Using the coarse-fine source model has already been shown effective for optimizing {\IVAconv} \cite{nakatani2021icassp}. We extend such use for updating the switching weights in this paper.

Modifying the optimization by introducing the coarse-fine source model is simple. We only need to calculate the source variances, $\lambda_{n,t}$, of the coarse source model in a step for updating separation matrices $\vect{W}^{(j)}_{f}$ by 
\begin{align}
    \lambda_{n,t} = \frac{1}{F}\sum_{f=1}^F\lambda_{n,t,f},\label{eq:fine2coarse}
\end{align}
and calculate Eq.~(\ref{eq:lambdaupdate}) using $\lambda_{n,f}$ substituted for $\lambda_{n,t,f}$.

\subsection{Processing flow of {\swIVAconv}}\label{sec:flow}
\begin{figure}
    \centering
    \includegraphics[width=\columnwidth]{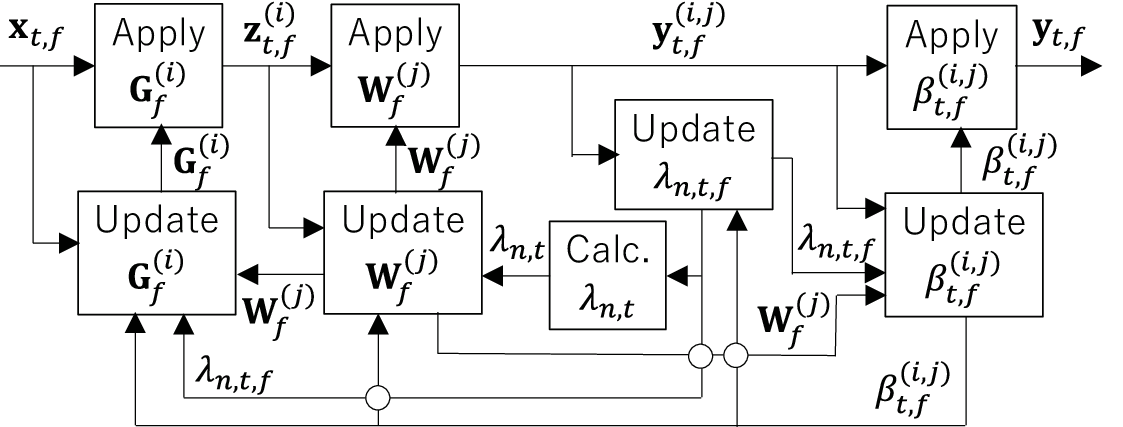}
    \caption{Schematic diagram of {\swIVAconv}: It also becomes schematic diagram of {\swIVA} when dropping two blocks on $\vect{G}_f$, directly putting $\vect{x}_{t,f}$ to two blocks on $\vect{W}_f$, and setting $I=1$.  {Small white circles at each intersection of lines denote that lines are not connected.}}
    \label{fig:flow}
\end{figure}

\SetKwInput{KwData}{Input}
\SetKwInput{KwResult}{Output}
\begin{algorithm}[t!]{
\SetAlgoLined
\DontPrintSemicolon
\AlgoDontDisplayBlockMarkers\SetAlgoNoEnd\SetAlgoNoLine
 \KwData{Observed signal $\{\vect{x}_{t,f}\}_{t,f}$\;}
 \KwResult{Estimated sources $\{y_{n,t,f}\}_{1\le n\le N,t,f}$}
 Initialize $\Lambda$, ${\cal B}$, and ${\cal W}$ by Algorithm 3\;
 \Repeat{convergence}{
  \tcc{swWPE part}
  $\vect{R}^{(i,j)}_{n,f}\leftarrow\sum_{t=1}^T\frac{\beta^{(i,j)}_{t,f}}{\lambda_{n,t,f}}\bar{\vect{x}}_{t,f}\bar{\vect{x}}_{t,f}^{\HT}$\, for $i,j,n,f$\;
  $\vect{P}^{(i,j)}_{n,f}\leftarrow\sum_{t=1}^T\frac{\beta^{(i,j)}_{t,f}}{\lambda_{n,t,f}}\bar{\vect{x}}_{t,f}\vect{x}_{t,f}^{\HT}$\, for $i,j,n,f$\;
  $\Psi^{(i)}_{f}\leftarrow{\displaystyle\sum_{j=1}^J\sum_{n=1}^M}\left(\vect{w}^{(j)}_{n,f}(\vect{w}^{(j)}_{n,f})^{\HT}\right)^*\otimes\vect{R}^{(i,j)}_{n,f}$\, for $i,f$\;
  $\Phi^{(i)}_{f}\leftarrow\sum_{j=1}^J\sum_{n=1}^M\vect{P}^{(i,j)}_{n,f}\left(\vect{w}^{(j)}_{n,f}(\vect{w}^{(j)}_{n,f})^{\HT}\right)$ for $i,f$\;
  $\mbox{vec}(\vect{G}^{(i)}_{f})\leftarrow(\Psi^{(i)}_{f})^{-1}\mbox{vec}(\Phi^{(i)}_{f})$\, for $i,f$\;
  $\vect{z}^{(i)}_{t,f}\leftarrow\vect{x}_{t,f}-(\vect{G}^{(i)}_{f})^{\HT}\bar{\vect{x}}_{t,f}$\, for $i,t,f$\;
  \tcc{swIVA part}
  Update ${\cal W}$, $\Lambda$, ${\cal B}$, and $\{\vect{y}_{t,f}\}_{t,f}$ by Algorithm 2\;
 }
 \caption{{\swIVAconv}.  
 }}
\end{algorithm}

\begin{algorithm}[t!]{
\SetAlgoLined
\DontPrintSemicolon
\AlgoDontDisplayBlockMarkers\SetAlgoNoEnd\SetAlgoNoLine
 \KwData{Updated dereverberated signal $\{\vect{z}^{(i)}_{t,f}\}_{i,t,f}$ and variables $\Lambda$, ${\cal B}$, and ${\cal W}$\;}
 \KwResult{Updated source estimates $\{\vect{y}_{t,f}\}_{t,f}$ and variables $\Lambda$, ${\cal B}$, and ${\cal W}$}
  \For{$K$ times}{
    $\lambda_{n,t} \leftarrow \frac{1}{F}\sum_{f=1}^F\lambda_{n,t,f}$\, for $n,t$\;
    $\Sigma^{(j)}_{n,f}\leftarrow\sum_{i=1}^I\sum_{t=1}^T\frac{\beta^{(i,j)}_{t,f}}{\lambda_{n,t}}\vect{z}^{(i)}_{t,f}(\vect{z}^{(i)}_{t,f})^{\HT}$\, for $j,n,f$\;
    $\vect{w}^{(j)}_{n,f}\leftarrow \left((\vect{W}^{(j)}_{f})^{\HT}\Sigma^{(j)}_{n,f}\right)^{-1}\vect{e}_n$\, for $j,n,f$\;
    $\vect{w}^{(j)}_{n,f}\leftarrow\vect{w}^{(j)}_{n,f}/\left((\vect{w}^{(j)}_{n,f})^{\HT}\Sigma^{(j)}_{n,f}\vect{w}^{(j)}_{n,f}\right)^{1/2}$\, for $j,n,f$\;
    $\vect{y}^{(i,j)}_{t,f}\leftarrow(\vect{W}^{(j)}_{f})^{\HT}\vect{z}^{(i)}_{t,f}$\, for $i,j,t,f$\;
    $\vect{y}_{t,f}\leftarrow\sum_{i=1}^I\sum_{j=1}^J\beta^{(i,j)}_{t,f}\vect{y}^{(i,j)}_{t,f}$\, for $t,f$\;
    $\lambda_{n,t,f}\leftarrow|y_{n,t,f}|^2+\varepsilon$\, for $n,t,f$\;
    Update $\{\beta_{t,f}^{(i,j)}\}_{i,j,t,f}$ using Eq.~(\ref{eq:betaupdate})
  }
 \caption{{\swIVA} update}}
\end{algorithm}

\begin{algorithm}[t!]{
\SetAlgoLined
\DontPrintSemicolon
\AlgoDontDisplayBlockMarkers\SetAlgoNoEnd\SetAlgoNoLine
 \KwData{Observed signal $\{\vect{x}_{t,f}\}_{t,f}$\;}
 \KwResult{Initialized $\Lambda$, ${\cal B}$ and ${\cal W}$}
 Initialize each element of $\{{\gamma}^{(i)}_{t,f}\}_{i,t,f}$ and $\{{\delta}^{(j)}_{t,f}\}_{j,t,f}$ at a random value in a range of $1\pm 10^{-3}$, and normalize it 
 to satisfy $\sum_{i=1}^I{\gamma}^{(i)}_{t,f}=1$ and $\sum_{j=1}^J{\delta}^{(j)}_{t,f}=1$\;
 \tcc{Iterate {\swWPE} one time\vspace{0.3mm}}
 $\lambda_{t,f}\leftarrow\vect{x}_{t,f}^{\HT}\vect{x}_{t,f}/M+\varepsilon$\, for $t,f$\;
 ${\vect{R}}^{(i)}_f\leftarrow\sum_t{\gamma}^{(i)}_{t,f}\bar{\vect{x}}_{t,f}\bar{\vect{x}}_{t,f}^{\HT}/{\lambda}_{t,f}$\, for $i,f$\;
 ${\vect{P}}^{(i)}_f\leftarrow\sum_t{\gamma}^{(i)}_{t,f}\bar{\vect{x}}_{t,f}{\vect{x}}_{t,f}^{\HT}/{\lambda}_{t,f}$\, for $i,f$\;
 ${\vect{G}}^{(i)}_f\leftarrow({\vect{R}}^{(i)}_f)^{-1}{\vect{P}}^{(i)}_f$\, for $i,f$\vspace{0.3mm}\;
 ${\vect{z}}^{(i)}_{t,f}\leftarrow\vect{x}_{t,f}-({\vect{G}}^{(i)}_{f})^{\HT}\bar{\vect{x}}_{t,f}$\, for $i,t,f$\vspace{0.3mm}\;
${\gamma}^{(i)}_{t,f}\leftarrow 
 \left\{\begin{array}{ll}
 1 & \mbox{if}~~i=\argmin_{i'}|{\vect{z}}^{(i')}_{t,f}|^2\\
 0 & \mbox{otherwise}
 \end{array}\right.$\, for $i,t,f$\;
 \tcc{Initialize {\swIVA} by Algorithm 2}
 $\vect{W}_f^{(j)}\leftarrow\vect{I}_M$\, for $j,f$\;
 $\lambda_{n,t,f}\leftarrow 1$\, for $n,t,f$\vspace{0.3mm}\;
 $\beta^{(i,j)}_{t,f}\leftarrow{\gamma}^{(i)}_{t,f}{\delta}^{(j)}_{t,f}$\, for $i,j,t,f$ 
 \vspace{0.3mm}\;
 Update ${\cal W}$, $\Lambda$, ${\cal B}$, and $\{\vect{y}_{t,f}\}_{t,f}$ by Algorithm 2\;
 \caption{Simple initialization}}
\end{algorithm}
Figure~\ref{fig:flow} shows a schematic diagram of {\swIVAconv} with the coarse-fine source model, and 
Algorithms 1, 2, and 3 summarize the processing flow implemented for our experiments. 

As in Algorithm~1, MCLP filters ${\cal G}$, separation matrices ${\cal W}$, source variances $\Lambda$, and switching weights ${\cal B}$ are updated by iterative optimization.  Lines~4 to 9 correspond to an update for {\swWPE} and line~10 corresponds to an update for {\swIVA}. Algorithm~2 shows more details of the flow of the update for {\swIVA}. Because the update for the separation matrices requires more iterations until convergence than for the MCLP filters {when using IP}, and the computational cost of the former is much smaller than the latter, we iterate the update of the separation matrices $K$ times in Algorithm~2 per update of the MCLP filter. {Preferable schemes for iterations depend on which algorithm we use for optimizing IVA.}

Algorithm~3 shows a simple initialization scheme that corresponds to one typically used by conventional {\IVAconv}. In it, after initializing the switching weights with random variables, we apply conventional {\swWPE} to obtain dereverberated observed signal $\{\vect{z}^{(i)}_{t,f}\}_{i,t,f}$ and update switching separation matrices $K$ times to obtain the initial values of ${\cal W}$, $\Lambda$, and ${\cal B}$.  The switching weights are first set as values between 0 and 1 and updated to binary values during the initialization steps.

\subsection{{\swIVA} formulation}\label{sec:swIVA}
This subsection briefly presents the formulation of {\swIVA} (not integrated with {\swWPE}) by summarizing its difference from the formulation of {\swIVAconv} as follows.
\begin{enumerate}
\item The observation model is assumed to exclude late reverberation, i.e., we set $\vect{l}_{n,t,f}=\vect{0}$ in Eq.~(\ref{eq:obs1}).
\item The same issues and motivations as in Sections~\ref{sec:conv} and \ref{sec:motivation} are applied except for the discussions on reverberation and dereverberation. 
\item The definition of a switching separation matrix used for {\swIVA} is obtained by dropping a switching MCLP filter from a swCBF, or more specifically by  Eqs.~(\ref{eq:yupdate}) and (\ref{eq:ysum}), substituting $\vect{x}_{t,f}$ for $\vect{z}^{(i)}_{t,f}$ and setting $I=1$.
\item The same assumptions as in Eqs.~(\ref{eq:independence}) and (\ref{eq:coarse}) are employed, and the same log-likelihood function as in Eqs.~(\ref{eq:swML1}) and (\ref{eq:swML2}) are derived except that we drop ${\cal G}$, substitute $\vect{x}_{t,f}$ for $\vect{z}^{(i)}_{t,f}$, and set $I=1$.
\item The optimization algorithm can be obtained from that of {\swIVAconv} by skipping the step that updates $\cal G$, substituting $\vect{x}_{t,f}$ for $\vect{z}^{(i)}_{t,f}$, and setting $I=1$. 
\item The same {\trick} is applied.
\end{enumerate}
As a result, the processing flow of {\swIVA} is obtained by modifying Algorithms~1, 2, and 3 as follows:
\begin{enumerate}
    \item All algorithms: substitute $\vect{x}_{t,f}$ for $\vect{z}^{(i)}_{t,f}$, and let $I=1$.
    \item Algorithm~1: skip lines~4 to 9.
    \item Algorithm~3: skip lines 2 to 8 and initialize $\beta^{(1,j)}_{t,f}$ at line 11 following the way for initializing ${\delta}^{(j)}_{t,f}$ at line 2.
\end{enumerate}
\section{Initialization techniques for {\swIVA}/{\swIVAconv}}\label{sec:spguide}
As already discussed in the introduction, the optimization of {\swIVA}/{\swIVAconv} is sensitive to initialization, i.e., it does not converge to good stationary points when using the simple initialization in Algorithm~3. Because there is no explicit constraint on the order of estimated sources, they can be easily permuted between different switching states especially at an early stage of the optimization. 
We call this the inter-state permutation problem.

To solve it, we designed two effective initialization schemes: 1) a blind single-state initialization and 2) a spatially guided initialization. The former allows us to perform fully blind optimization, and the latter enables us to utilize spatial information obtained, e.g., by NN-based TF mask estimation.

\subsection{Blind single-state initialization}

With blind single-state initialization, we use the {\swIVA}/{\swIVAconv} themselves for the initialization, but tentatively setting the number of switching states $J=1$, i.e., using only one separation matrix in {\swIVA}/{\swIVAconv}. Because {\swIVA}/{\swIVAconv} with $J=1$ are free from the inter-state permutation problem, we can avoid such permutation errors. Note that {\swIVA} with $J=1$ is equivalent to {\IVA}.

We implemented it as follows:
\begin{enumerate}
\item We started and iterated the optimization of {\swIVA}/{\swIVAconv} by setting $J=1$ and following Algorithms~1 to 3. 
\item After a certain predetermined number of iterations, we reset the number of switching states $J$ to its original number specified for the estimation and re-initialized {\swIVA}/{\swIVAconv} based on their updated parameters estimated with $J=1$.
\item We continued the remaining iterations for the optimization.
\end{enumerate}
In our experiment, we performed the following re-initialization in the above step~2 before line~4 of Algorithm~2:
\begin{enumerate}
    \item Copy updated separation matrix $\vect{W}^{(1)}_{t,f}$ to re-initialized separation matrices $\vect{W}^{(j)}_{t,f}$ for $1\le j\le J$.
    \item Initialize ${\delta}^{(j)}_{t,f}$ at random values following line~2 of Algorithm~3, and multiply it with updated switching weights $\beta^{(i,1)}_{t,f}$ to obtain re-initialized switching weights $\beta^{(i,j)}_{t,f}$ for all $i$, $j$, $t$, and $f$ as
    \begin{align}
        \beta^{(i,j)}_{t,f}\leftarrow\beta^{(i,1)}_{t,f}{\delta}^{(j)}_{t,f}.
    \end{align}
\end{enumerate}

\subsection{Spatially guided initialization}
With this initialization, we use the ATFs of the source signals separately estimated from the observed signal. Such ATFs can be reliably estimated for conventional mask-based BFs/CBFs, e.g., with a TF-mask estimation based on a neural network (NN) \cite{nakatani2020taslp} or blind spatial clustering \cite{GSS,ito16eusipco}. So we adopt such a technique in this paper. 

In the following, we first explain how ATFs can be estimated based on TF masks and then describe how to initialize {\swIVA}/{\swIVAconv} using the estimated ATFs. (See Section~\ref{sec:expset} where we estimated the TF masks in our experiments.)

\subsubsection{TF mask-based ATF estimation}\label{sec:ATF}
We adopted an ATF estimation method based on generalized eigenvalue decomposition \cite{MGolan09taslp,Ziteng2018SPCom}. 
During the initialization step in Algorithm~3, dereverberated observed signal ${\vect{z}}_{t,f}$ can be obtained based on {\swWPE} after line~7 of Algorithm~3:
\begin{align}
    {\vect{z}}_{t,f}=\sum_{i=1}^I{\gamma}^{(i)}_{t,f}{\vect{z}}^{(i)}_{t,f}.
\end{align}
For {\swIVA}, we substituted $\vect{x}_{t,f}$ for $\vect{z}_{t,f}$ and set $I=1$.
Then, we estimated TF masks $\Omega_{n,t,f}$ from ${\vect{z}}_{t,f}$, based, e.g., on a neural network. $\Omega_{n,t,f}$ takes a value between 0 and 1, where $\Omega_{n,t,f}=1$ and 0 respectively mean that the TF point is dominated by the $n$th source and some other sources.  With the masks, the covariance matrix of the $n$th source, denoted by $\Gamma_{Z,n,f}$, and that of interference signals, denoted by $\Gamma_{V,n,f}$, are calculated:
\begin{align}
    \Gamma_{Z,n,f}&=\frac{\sum_{t=1}^{T}\Omega_{n,t,f}{\vect{z}}_{t,f}{\vect{z}}_{t,f}^{\HT}}{\sum_{t=1}^{T}\Omega_{n,t,f}},\\
    \Gamma_{V,n,f}&=\frac{\sum_{t=1}^{T}(1-\Omega_{n,t,f}){\vect{z}}_{t,f}{\vect{z}}_{t,f}^{\HT}}{\sum_{t=1}^{T}(1-\Omega_{n,t,f})},
\end{align}
Then the ATFs can be estimated:
\begin{align}
    \vect{a}_{n,f}&\leftarrow\Gamma_{V,n,f}\mbox{MaxEig}(\Gamma_{V,n,f}^{-1}\Gamma_{Z,n,f}),
\end{align}
where $\mbox{MaxEig}(\cdot)$ is a function that extracts the eigenvector corresponding to the maximum eigenvalue. 

\subsubsection{Initialization using estimated ATFs}\label{sec:sginit}
This technique provides a way to initialize separation matrices $\cal W$ and source variances $\Lambda$ as a substitution of lines~9 and 10 in Algorithm~3. First, we employed the conventional Minimum Power Distortionless Response (MPDR) BF for the initialization of ${\cal W}$, where the estimated ATFs are used in the distortionless constraint,  $(\vect{w}^{(j)}_{n,f})^{\HT}\vect{a}_{n,f}=a_{n,r,f}$, where $a_{n,r,f}$ is the reference channel element of $\vect{a}_{n,f}$. 
In concrete, each column $\vect{w}^{(j)}_{n,f}$ of $\vect{W}^{(j)}_{f}$ for $1\le n\le N$ and all $j$ and $f$ is initialized:
\begin{align}
    \vect{w}^{(j)}_{n,f}&=\frac{a_{n,r,f}^*(\Xi^{(j)}_{f})^{-1}\vect{a}_{n,f}}{(\vect{a}_{n,f})^{\HT}(\Xi^{(j)}_{f})^{-1}\vect{a}_{n,f}},
\end{align}
where {$\Xi_{f}^{(j)}$ is a weighted covariance matrix of $\mathbf{z}_{t,f}$, calculated with the switching weights $\delta_{t,f}^{(j)}$
initialized at line~2 of Algorithm~3:}
\begin{align}
    \Xi^{(j)}_{f}&=\frac{\sum_{t=1}^{T}\delta^{(j)}_{t,f}{\vect{z}}_{t,f}({\vect{z}}_{t,f})^{\HT}}{\sum_{t=1}^{T}\delta^{(j)}_{t,f}}.\label{eq:initcbf2}
\end{align}
As for filter coefficients to separate the noise components, i.e., $\vect{w}^{(j)}_{n,f}$ for $N+1\le n\le M$, we initialized them as $\vect{w}^{(j)}_{n,f}=\vect{e}_n$. 

The variance of each source is then initialized as the squared average of $J$ BFs' outputs:
\begin{align}
    \lambda_{n,t,f}=\left|\frac{\sum_{j=1}^J\delta^{(j)}_{t,f}(\vect{w}^{(j)}_{n,f})^{\HT}{\vect{z}}_{t,f}}{\sum_{j=1}^J\delta^{(j)}_{t,f}}\right|^2.\label{eq:init3}
\end{align}

\section{{Computational time complexity}}
\begin{table}[t]
\centering
\setlength{\tabcolsep}{5pt}
{
\caption{Computational time complexity per iteration of covariance matrix calculation (Cov), matrix inversion (Inv), filtering (Filtering), and switching weight update (Switch) with 
$M$:~\#Microphones, $F$: \#Frequencies, $T$: \#Frames, $L$: Length \\of a CBF, and $(I,J)$: \# of switching states for WPE and IVA.}\label{tab:complexity}
\begin{tabular}{c|cc}\toprule
& IVA & swIVA \\\hline
\rule[0mm]{0mm}{3mm}Cov & $O(M^3TF)$ & $O(M^3TF)$\\
Inv & $ O(M^4F)$ & $O(JM^4F)$\\
Filtering & $O(M^2TF)$ & $O(JM^2TF)$\\
Switch & - & $O(JMTF+JM^3F)$ \\\hline
\rule[0mm]{0mm}{3mm}& CIVA & swCIVA \\\hline
\rule[0mm]{0mm}{3mm}Cov & $O(M^3L^2TF)$ & $O(M^3L^2TF+IJM^5L^2F)$\\
Inv & $O(M^4L^3F)$ & $O(IM^6L^3F+JM^4F)$\\
Filtering & $O(M^2LTF+M^2TF)$ & $O(IM^2LTF+IJM^2TF)$\\
Switch & - & $O(IJMTF+JM^3F)$ \\
\bottomrule
\end{tabular}}
\end{table}
{
Table~\ref{tab:complexity} summarizes the computational time complexity per iteration of the major processing blocks for IVA, swIVA, CIVA, and swCIVA. First, swIVA and IVA have the same complexity for covariance matrix calculation (Cov) in line~4 of Algorithm~2 
because $\beta_{t,f}^{(i,j)}$ with $I=1$ and $J>1$ can take a value 1 for only one of states $j$ at each TF point, and $\vect{z}_{t,f}^{(i)}(\vect{z}_{t,f}^{(i)})^{\HT}$ ($=\vect{x}_{t,f}\vect{x}_{t,f}^{\HT}$) is calculated only once in each iteration. In contrast, for matrix inversion (Inv) in line~5 of Algorithm~2 
and filtering (Filtering) in line~7 of Algorithm~2, 
swIVA needs to conduct the same processing for each state $j$, and thus it has $J$ times larger complexity than IVA.}

{As for {\swIVAconv} and {\IVAconv}, the table shows the complexity using source-wise covariance decomposition for {\swIVAconv} and source-wise CBF factorization \cite{nakatani2020interspeech} for {\IVAconv}.  Thus, the complexity of {\swIVAconv} increases not only by the use of the switching mechanism but also by the source-wise covariance decomposition. The complexity of {\swIVAconv} has an extra term $O(IJM^5L^2F)$ for Cov in line~6 of Algorithm~1, 
and is $IM^2$ times a higher order for Inv in line~8 of Algorithm~1. For Filtering, {\swIVAconv} has $I$ times more complexity in line~9 of Algorithm~1 
and $IJ$ times more complexity in line~7 of Algorithm~2. 
}

{For updating the switching weights (Switch) in line~10 of Algorithm~2, 
the complexity for calculating the first line of Eq.~(\ref{eq:swML2}) is $O(JMTF)$ for {\swIVA} and $O(IJMTF)$ for {\swIVAconv}, which are $M$ times smaller than that of Filtering for {\swIVA} and for {\swIVAconv}. 
The complexity for calculating the second line, i.e., $\det\vect{W}_f^{(j)}$, is $O(JM^3F)$, which is $M$ times smaller than that of Inv for {\swIVA} and for {\swIVAconv}.}

{In summary, when $T\gg JM$ and $M\gg J$, the computational cost for {\swIVA} is dominated by Cov with $O(M^3TF)$, which is the same as that for {\IVA}. Similarly, when $T\gg IJM^2+IM^3L$, $ML\gg I$, and $ML^2\gg IJ$ for {\swIVAconv}, the computational cost is dominated by Cov with $O(M^3L^2TF)$, which is the same as that for {\IVAconv}. As $J$ increases for {\swIVA} and as $I$, $J$, and $M$ increase for {\swIVAconv}, the costs become significantly larger than those of the conventional methods.}

\section{Experiments}
\label{sec:experiments}

This section experimentally evaluates {\swIVA}/{\swIVAconv} in terms of signal distortion reduction and ASR improvement.  Tables~\ref{tab:methods} and \ref{tab:configs} summarize the methods and configurations compared in the experiments.
First, we confirmed the necessity of the proposed initialization techniques and the coarse-fine source model. Then we evaluated the joint optimization in comparison with separate optimization, for which we applied conventional {\swWPE} \cite{SWWPE2021} and {\swIVA} in a cascade configuration without interaction. Finally, we evaluated {\swIVA}/{\swIVAconv} with varying numbers of switching states.

\subsection{Dataset, evaluation metrics, and analysis condition}\label{sec:expset}

\begin{table}[t]
    \centering
    \setlength{\tabcolsep}{3pt}
    \caption{Methods to be compared}    \label{tab:methods}
    \begin{tabular}{ccccc}\toprule
    Method &~& Switch &~& \#States for switch(es)\\\cline{1-1}\cline{3-3}\cline{5-5}
    \rule[0mm]{0mm}{3mm}{\IVA} \cite{IP} && - && $J=1$\\
    {\swIVA} && $\checkmark$ && $J=2$ or $3$\\
    {\IVAconv} \cite{ikeshita2019waspaa,nakatani2020interspeech} && - && $(I,J)=(1,1)$\\
    {\swIVAconv} && $\checkmark$ && $(I,J)=(1,2),(2,1),(2,2),\ldots$, or $(3,3)$\\\bottomrule
    \end{tabular}\bigskip
\setlength{\tabcolsep}{4pt}
    \caption{Configurations to be compared: `*' indicates a default configuration used in experiments unless otherwise noted.}\label{tab:configs}
    \begin{tabular}{lcl}\toprule
    Parameter &~& Configurations \\\cline{1-1}\cline{3-3}
    \rule[0mm]{0mm}{3mm}Initialization && Simple, blind single-state, or *spatially guided\\
    Source model && Switch: Coarse (C) or *Fine (F), MCLP: C or *F\\
    Optimization && *Joint or separate\\
    Switching model && direct (Fig.~1(a)) or *Factorized (Fig.~1(b))\\\bottomrule
    \end{tabular}
\end{table}
To evaluate the estimated source signals, we used {two different datasets}: REVERB-2MIX \cite{reverb2mix} {and TIMIT-ConvMix. REVERB-2MIX was used in Sections~\ref{results} to \ref{sec:varstates} and its specifications are described in the following. TIMIT-ConvMix was used only in Section~\ref{sec:TIMIT}, where we also summarized its specifications.}

REVERB-2MIX is composed of noisy reverberant speech mixtures. Each mixture in the dataset was created by mixing two utterances (i.e., $N=2$), extracted from the REVERB Challenge dataset (REVERB) \cite{REVERB}: one from its development set (Dev set) and another from its evaluation set (Eval set). REVERB-2MIX is composed of two subsets, SimuData and RealData, created from SimuData and RealData of REVERB. {Table~\ref{tab:my_label} shows some statistics of each subset.} REVERB SimuData was generated by convolving the measured impulse responses with utterances extracted from the WSJCAM0 dataset \cite{wsjcam0} and by adding measured stationary noise. The reverberant signal-to-noise ratio (SNR) was set to 20 dB, {and reverberation time RT60 was varied from 0.2 to 0.7 s}. REVERB RealData was recorded in actual noisy reverberant rooms, where WSJCAM0 transcriptions were read by humans. Although 8-channel data are available for REVERB-2MIX, we only used its first 2 or 3 channels in the experiments {and estimated the same number of sources as channels in the respective settings}. Following the REVERB-2MIX guidelines, an evaluation was performed using separated signals that correspond to the Eval set of REVERB, selected from the estimated speech signals based on the correlation with the original signals in the REVERB Eval set.

\begin{table}[t]
    \centering
    \setlength{\tabcolsep}{3.7pt}
{
    \caption{Statistics of mixed utterances in REVERB-2MIX dataset}
    \label{tab:my_label}
    \begin{tabular}{c|cccccccc}\toprule
         & && \multicolumn{2}{c}{Each mixture} && \multicolumn{3}{c}{Length of mixtures}\\\cline{4-5}\cline{7-9}
         &  \#Mixtures && \#Sources & \#Mics. && Average & Min & Max \\\hline
    \rule[0mm]{0mm}{3mm}SimuData & 2176 && 2 & 2 or 3 && 7.9 s & 2.9 s & 15.6 s\\
    RealData & 372 && 2 & 2 or 3 && 6.5 s & 1.9 s & 14.5 s\\\bottomrule
    \end{tabular}
}
\end{table}

We used the REVERB-2MIX SimuData to evaluate the distortions of the enhanced signals and the RealData to evaluate the ASR performance of the enhanced signals, {where all the mixtures in the respective subsets were used for the evaluations.}
As a metric to evaluate the signal distortions \cite{metrics}, we adopted the Frequency-Weighted Segmental SNR (FWSSNR), which was shown in the REVERB challenge \cite{REVERB} to have relatively high correlation with a subjective evaluation on the perceived amount of reverberation. To evaluate the ASR performance, we calculated the Word Error Rates (WERs) of the enhanced signals using a baseline ASR system developed for REVERB using Kaldi \cite{kaldi}. This system adopted a Time-Delay NN (TDNN) acoustic model trained using Lattice-Free Maximum Mutual Information (LF-MMI) and online i-vector extraction and a trigram language model. They were all trained on the REVERB training set. 

We set the frame length and the shift to 32 and 8 ms and used a Hann window for a short-time analysis. The sampling frequency was 16 kHz. For an MCLP filter, the prediction delay was set at $D = 2$ and the {CBF} length was set at $L = 10$. For {\swIVAconv}, separation matrices were updated $K=5$ times per MCLP filter update, and the separation matrices and the MCLP filters were updated 50 and 10 times. 
We used diagonal loading for the matrix inversions to make them stable in the optimization.
{For example, covariance matrices calculated for updating the filter coefficients of WPE and IVA become rank deficient when the number of active frames for a certain state at a frequency is smaller than the sizes of the matrices. We set the diagonal loading factors at $10^{-4}\times\mbox{Trace}({\Psi}_f^{(i)})$ for WPE in Eqs.~(\ref{eq:gupdate}) and $10^{-10}\times\mbox{Trace}((\vect{W}_f^{(j)})^{\HT}\Sigma_{n,f}^{(j)})$ for IVE in Eq.~(\ref{eq:wupdate1}).
}
For all the methods, we applied projection back \cite{projback} post-processing to solve the scale ambiguity of BSS.
For the estimation of the time-frequency masks, which were used to estimate the ATFs for the spatially guided initialization, we adopted a frequency-domain convolutional NN (CNN) \cite{CNN-uPIT} with a large receptive field, similar to the one used in a fully-Convolutional Time-domain Audio Separation Network (Conv-TasNet) \cite{ConvTasnet2019TASLP}. We trained the CNN to estimate masks based on utterance-level permutation invariant training \cite{pit17taslp}. 

\subsection{Importance of initialization}
\label{results}

\newcommand{\figind}{\hspace{6.5mm}}
\begin{figure}
\includegraphics[width=\columnwidth]{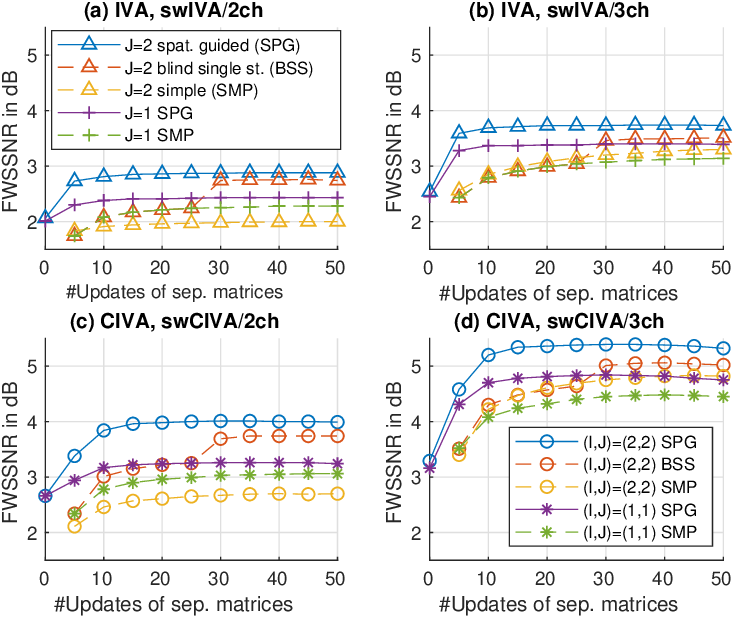}
\caption{FWSSNRs in dB obtained by {\IVA} ($J=1$), {\swIVA} ($J=2$), {\IVAconv} ($(I,J)=(1,1)$), and {\swIVAconv} ($(I,J)=(2,2)$) with spatially guided initialization (SPG), blind single-state initialization (BSS), or simple initialization (SMP): Horizontal axis indicates \#updates of separation matrices. \#Updates 0 and 5 correspond to MPDR BF outputs used in spatially guided initialization and {\swIVA} outputs obtained after 5 updates in Algorithm~3. For blind single-state initialization, we used first 25 updates as the initialization steps.
Default configurations in Table~\ref{tab:configs} were used for other parameters.}\label{exp:init}
\end{figure}

In the first experiment, we compared the performances of {\IVA}, {\IVAconv}, {\swIVA} and {\swIVAconv} using three types of initialization techniques, simple initialization in Algorithm~3, blind single-state initialization, and spatially guided initialization. Figure~\ref{exp:init} shows the FWSSNRs of the enhanced signals. 

First, comparing the results with {\swIVA} ($J=2$) and {\IVA} ($J=1$) when using the spatially guided initialization, {\swIVA} greatly outperformed {\IVA} in both (a) and (b).
{Also, {\swIVA} with blind single-state initialization shows a clear performance jump between 25 and 30 updates of separation matrices when the single-state initialization steps finished, and in the following updates it substantially outperformed {\IVA} and {\IVAconv} using the simple initialization.}
In contrast, {\swIVA} underperformed {\IVA} with the simple initialization for the 2-ch case.
Next, comparing the results using {\swIVA} ($J=2$), the spatially guided initialization largely outperformed the others, and the blind single-state initialization was second. Note that spatially guided initialization was also very effective for {\IVA} ($J=1$). 
As for {\swIVAconv}, we confirmed the identical tendency in (c) and (d).

These results clearly show the importance of the initialization of {\swIVA} and {\swIVAconv}. 
Without appropriate initialization, the switching mechanism might degrade the performance. With our proposed initialization techniques, in contrast, the switching mechanism substantially and consistently improved the performance more than without it. {These results also suggest that the permutation alignment was robustly kept once it was appropriately initialized although it was not guaranteed.} 
The blind single-state initialization helped {\swIVA} and {\swIVAconv} work effectively based only on blind processing, and the spatially guided initialization further improved performance based on NN-based mask estimation. 

Below we used spatially guided initialization as a default configuration in Table~\ref{tab:configs}.

\subsection{Necessity of coarse-fine source model}

\begin{figure}
\includegraphics[width=\columnwidth]{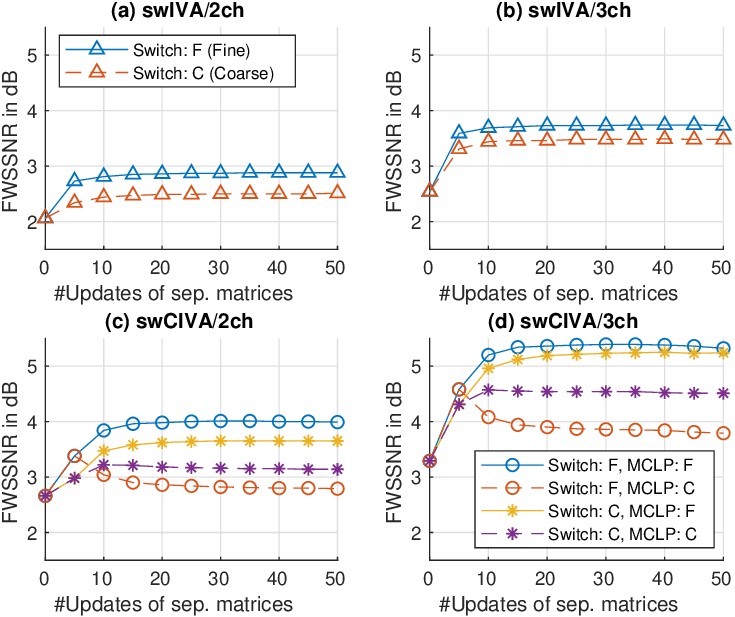}
\caption{FWSSNRs in dB obtained with different source models, indicated as C (Coarse) and F (Fine), for optimization of switches (Switch) and MCLP filters (MCLP) using {\swIVA} ($J=2$) and {\swIVAconv} ($(I,J)=(2,2))$: Default configurations in Table~\ref{tab:configs} are used for other parameters.}\label{exp:CF}
\end{figure}
\begin{figure}
\includegraphics[width=\columnwidth]{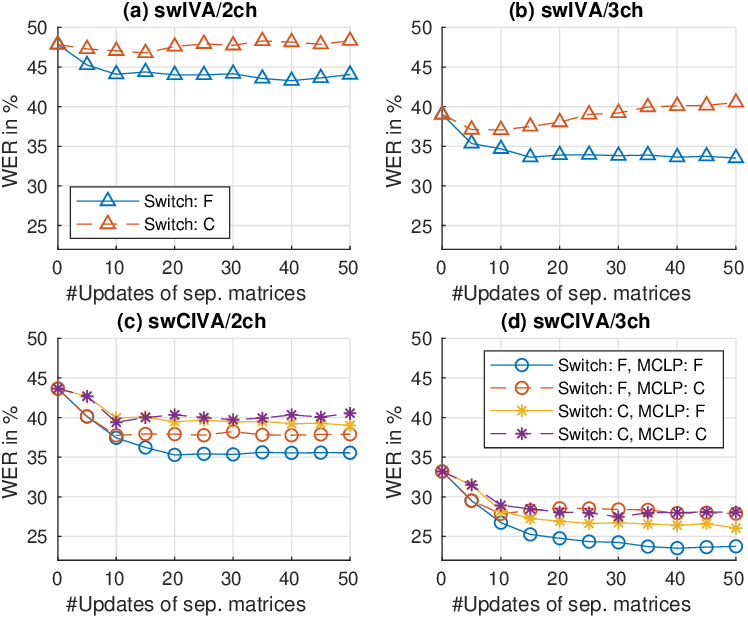}
\caption{WERs in \% obtained with different source models, indicated as C (Coarse) and F (Fine), for optimization of switches (Switch) and MCLP filters (MCLP) using {\swIVA} ($J=2$) and {\swIVAconv} ($(I,J)=(2,2))$: Default configurations in Table~\ref{tab:configs} are used for other parameters.}\label{exp:CFWER}
\end{figure}

Figures~\ref{exp:CF} and \ref{exp:CFWER} show the FWSSNRs and WERs of the signals enhanced using {\swIVA}/{\swIVAconv} by modifying the configuration of the source model between coarse and fine models.  We set $J=2$ for {\swIVA} and $(I,J)=(2,2)$ for {\swIVAconv}.  In the figure, ``Switch: F'' and ``MCLP: F'' mean that the fine source model was used for switching weights and MCLP filters. ``Switch: C'' and ``MCLP: C'' mean that the coarse model was used instead. For the separation matrix, we used the coarse model in all cases. 

The figures show that using the fine source model for both the switching weights and the MCLP filters greatly improved the performance of {\swIVA}/{\swIVAconv} for all cases especially in comparison when the coarse model was used for both of them. 

These results clearly show the necessity of respectively using coarse and fine models for the optimization of {\swIVA} and for that of the MCLP filters and switches. 

\begin{figure}
\includegraphics[width=\columnwidth]{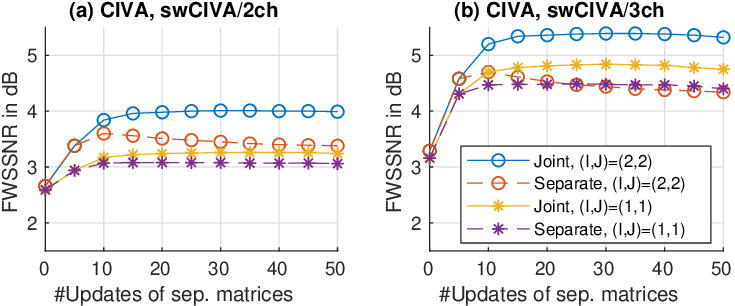}\smallskip\\
\includegraphics[width=\columnwidth]{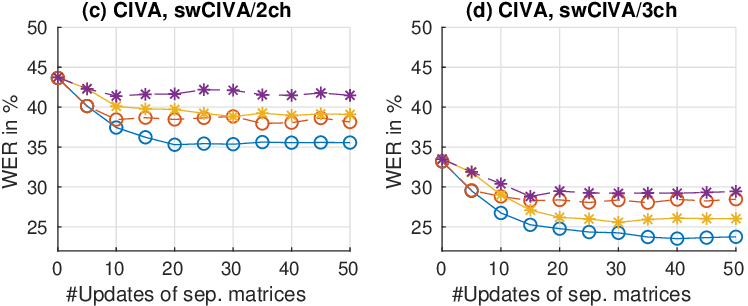}
\caption{FWSSNRs in dB and WERs in \% obtained by {\IVAconv} ($(I,J)=(1,1)$) and {\swIVAconv} $(I,J)=(2,2)$ using joint optimization (Joint) and separate optimization (Separate): Default configurations in Table~\ref{tab:configs} are used for other parameters.}\label{exp:Joint}
\end{figure}

{Note that the FWSSNRs and WERs obtained by {\swIVA} and {\swIVAconv} did not necessarily improve monotonically as the numbers of updates were increased even using the preferable configurations. For example, see the blue line after 30 updates in Fig.~\ref{exp:CF}(d) and the blue line after 40 updates in Fig.~\ref{exp:CFWER}(a). 
This is probably because the optimization was performed based not on FWSSNRs or WERs but on the ML criterion. The performance degradation might be caused by over-fitting to the ML criterion.} 

\subsection{Joint optimization vs separate optimization}

Next, we evaluated the effect of the joint optimization of {\IVAconv} ($(I,J)=(1,1)$) and {\swIVAconv} ($(I,J)=(2,2)$) in comparison with the case using separate optimization. With separate optimization, we connected conventional {\swWPE} \cite{SWWPE2021} and {\swIVA} in a cascade configuration and optimized each separately without interaction between {\swIVA} to {\swWPE} except for passing the output of {\swWPE} to {\swIVA}. Overall optimality is not guaranteed for separate optimization. 

Figure~\ref{exp:Joint} shows the results using 2-ch/3-ch observations. In all cases, the joint optimization largely outperformed the separate optimization. The joint optimization with {\IVAconv} ($(I,J)=(1,1)$) even outperformed the separate optimization with {\swIVAconv} ($(I,J)=(2,2)$) for a 3-ch observation.

This result shows that joint optimization is essential for {\IVAconv} and {\swIVAconv}. 

\begin{figure}
\includegraphics[width=\columnwidth]{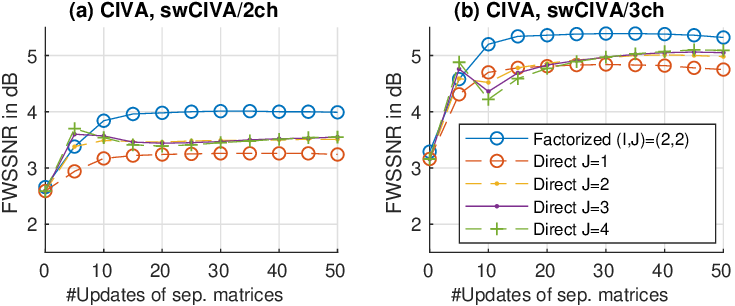}
\caption{FWSSNRs in dB obtained by {\IVAconv} and {\swIVAconv} using factorized switching model $(I,J)=(2,2)$ and direct switching model ($J=1,2,3$ and $4$): Default configurations in Table~\ref{tab:configs} are used for other parameters.}
\label{fig:vsdirect}
\end{figure}

\begin{figure}
\includegraphics[width=\columnwidth]{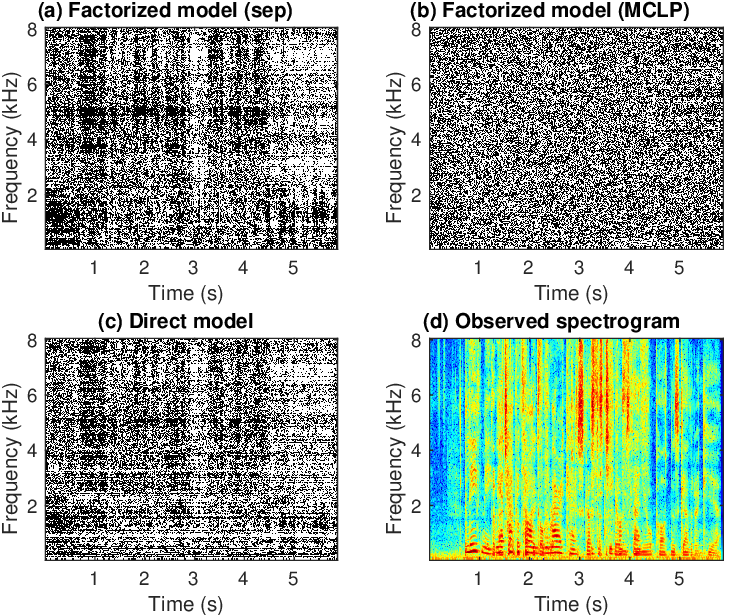}\smallskip\\
\includegraphics[width=8.63cm]{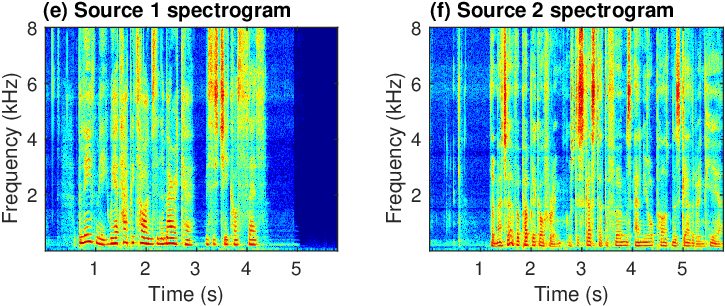}
\caption{Switching weights of (a) separation matrix and (b) MCLP filter estimated with factorized switching model ($(I,J)=(2,2)$), and (c) those estimated with direct switching model ($J=2$) when using 2-ch observation: Black and white at each TF point respectively represent switching weight value: 0 and 1. Each figure shows only one of switching weights estimated for each switch. With factorized model, switching weights of (a) separation matrix and (b) MCLP filter are respectively calculated by $\sum_{i=1}^{I}\beta^{(i,j)}_{t,f}$ and $\sum_{j=1}^J\beta^{(i,j)}_{t,f}$. Spectrograms of (d) observed mixture, (e) source 1, and (f) source 2 are also shown. Default configurations in Table~\ref{tab:configs} are used for other parameters.}    \label{fig:switches}
\end{figure}

\subsection{Factorized switching model vs direct switching model}\label{exp:switches}

Figure~\ref{fig:vsdirect} compares FWSSNRs obtained by {\swIVAconv} based on the direct and factorized switching models (Figs.~1(a) and (b)) using 2-ch and 3-ch microphones. The same processing flow shown in Algorithms~1 to 3 was used with the direct switching model except that the same switch was used for both the MCLP filters and the separation matrices. The figure shows that the factorized model ($(I,J)=(2,2)$) was substantially better than the direct model. Although the latter improved the performance when we increased the number of switching states from $J=1$ (i.e., conventional {\IVAconv}) to $2$, further increasing the number did not improve the performance.

Figure~\ref{fig:switches} shows switching weights estimated by {\swIVAconv} with the factorized switching model ($(I,J)=(2,2)$) and the direct switching model ($J=2$). 
A spectrogram of the observed mixture is also shown. 
We can clearly see different patterns in (a) and (b), i.e., switches estimated for separation matrices and MCLP filters with the factorial switching model. In contrast, the switches estimated with the direct switching model have a pattern relatively close to (a), although slightly resembling a mixture of (a) and (b). Note that (a) seems closely related to (d), which is the observed spectrogram structure, but we can hardly see such a relationship between (b) and (d).

These results suggest that the two types of switches in {\swIVAconv}, one for MCLP filters and the other for separation matrices, capture rather different time-varying characteristics of the observed signal. 
{This result might be caused by the different roles played by the MCLP filters and separation matrices. The former reduces the late reverberation of all the sources and the latter reduces the direct and early reflection components of the extraneous speaker and stationary noise. Because the time-varying characteristics of these interference signals are completely different from each other, the TF patterns of their switches might become very different.}
These results show the importance of the factorized switching model for making {\swIVAconv} work effectively. 

\begin{figure}
\includegraphics[width=\columnwidth]{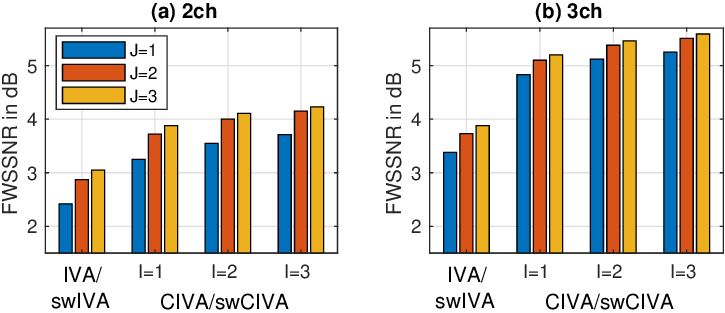}
\caption{FWSSNRs in dB obtained with different number of switching states  using {\IVA} ($J=1$), {\swIVA} ($J=2$ or 3), {\IVAconv} ($(I,J)=(1,1)$), and  {\swIVAconv} ($(I,J)=(1,2),(2,1),(2,2),\ldots$, or $(3,3)$) after 25 updates: Default configurations in Table~\ref{tab:configs} are used for other parameters.}\label{exp:switch_snr}
\end{figure}

\begin{figure}
\includegraphics[width=\columnwidth]{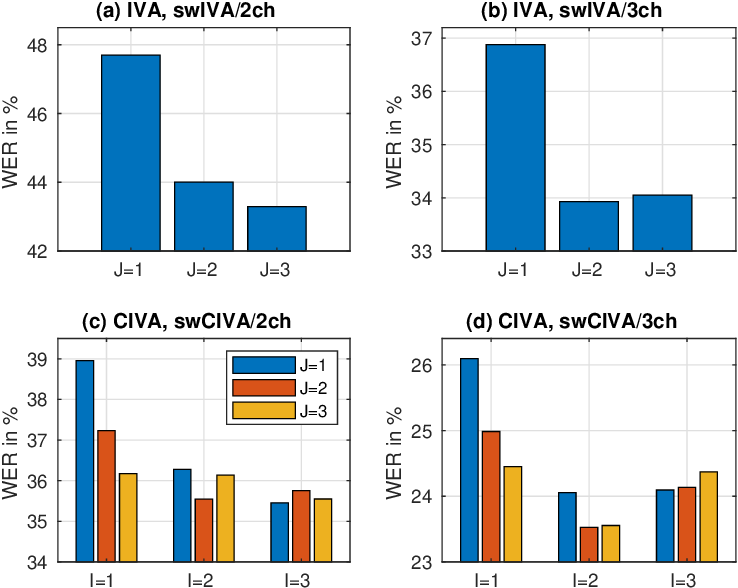}
\caption{WERs in \% obtained with different number of switching states  using {\IVA} ($J=1$), {\IVAconv} ($(I,J)=(1,1)$), {\swIVA} ($J=2$ or 3), and  {\swIVAconv} ($(I,J)=(1,2),(2,1),(2,2),\ldots$, or $(3,3)$) after 40 updates: Default configurations in Table~\ref{tab:configs} are used for other parameters. $Y$-axes are individually adjusted for making the differences in each plot clear. }\label{exp:switch_wer}
\end{figure}

\subsection{Evaluation with various numbers of switching states}\label{sec:varstates}

In this experiment, we varied the number of switching states for {\swIVA} as $J=2$ or $3$ and those for {\swIVAconv} as $(I,J)=(1,2),(2,1),(2,2),\ldots,(3,3)$ to examine their effect on FWSSNRs and WERs. {\IVA} ($J=1$) and {\IVAconv} ($(I,J)=(1,1)$) were also compared.

Figure~\ref{exp:switch_snr} shows the FWSSNRs obtained after 25 updates of the separation matrices. The FWSSNRs were consistently improved by respectively increasing $I$ and $J$ from 1 to 3. On the other hand, Fig.~\ref{exp:switch_wer} shows the WERs obtained after 40 updates of the separation matrices. Because the WER convergences were slightly slower than those of FWSSNRs, the WERs obtained at a slightly later iteration step were shown here. The WERs were improved by increasing $I$ and $J$ from 1 to 2, but a further increase did not necessarily improve them; it sometimes even degraded them. This may be caused by overfitting {\swIVA}/{\swIVAconv} due to their large numbers of parameters linearly increasing as $I$ and $J$ increase. 

These results show that {\swIVA} and {\swIVAconv} effectively reduced the signal distortions and improved the ASR performance, although it became unstable in terms of ASR when increasing the number of switching states to 3.

\subsection{Evaluation using a different dataset: TIMIT-ConvMix}\label{sec:TIMIT}

In this experiment, we evaluated our proposed methods using TIMIT-ConvMix, which is composed of simulated noisy reverberant mixtures. Table~\ref{tab:TIMIT} summarizes the statistics of the dataset. To generate the mixtures, we first concatenated utterances extracted from the TIMIT corpus \cite{TIMIT} to obtain a set of single speaker clean utterance sequences whose length was at least 10~s. Then we mixed two or three utterance sequences and five different additive noise signals extracted from the CHiME-3 dataset \cite{chime3} after individually reverberating them. Room impulse responses (RIRs) extracted from JR1 in the RWCP dataset \cite{RWCP} were used. Its RT60 was 0.6~s. All four types of noise signals were used, BUS, STR, PED, and CAF, although each mixture contained only a single type. We set the power ratio of each reverberant speech signal to the sum of the additive noise signals to 10~dB.

The major differences of this dataset from REVERB-2MIX are that all speakers continuously speak during whole sequences under a higher noise level condition. Also, we can examine mixtures with three sources. In this evaluation, we adopted signal-to-distortion ratio (SDR) as the evaluation metric \cite{BSSEVAL}, which is widely used in source separation research. We used the MUSEVAL V4 toolkit \cite{museval} with its bss\_eval\_images configuration. As reference signals, we used clean utterance sequences that were convolved with the initial 32 ms part of the RIRs used for generating the corresponding mixtures.  To solve the permutation ambiguity between the enhanced sources, we selected a combination of sources that gave the best average SDR score for each mixture.

\begin{table}[t]
    \centering
    \setlength{\tabcolsep}{4.5pt}
{
    \caption{Statistics of mixed utterances in TIMIT-ConvMix dataset}
    \label{tab:TIMIT}
    \begin{tabular}{c|ccccccc}\toprule
         & & \multicolumn{2}{c}{Each mixture} && \multicolumn{3}{c}{Length of mixtures}\\\cline{3-4}\cline{6-8}
         &  \#Mixtures & \#Sources & \#Mics. && Average & Min & Max \\\hline
    \rule[0mm]{0mm}{3mm}2-Mix & 40 & 2 & 3 && 12.3 s & 11.1 s & 13.6 s\\
    3-Mix & 40 & 3 & 4 && 12.6 s & 12.0 s & 13.6 s\\\bottomrule
    \end{tabular}\bigskip
}
\end{table}

\begin{figure}
\includegraphics[width=\columnwidth]{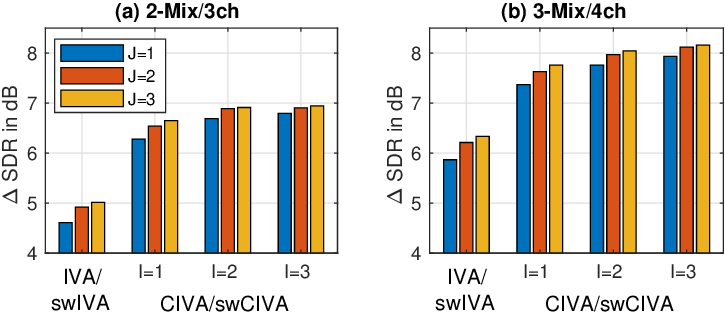}
\caption{{Improvement of SDRs in dB on 2-Mix and 3-Mix of TIMIT-ConvMix using {\IVA} ($J=1$), {\IVAconv} ($(I,J)=(1,1)$), {\swIVA} ($J=2$ or 3), and  {\swIVAconv} ($(I,J)=(1,2),(2,1),(2,2),\ldots$, or $(3,3)$) after 50 updates of separation matrices: Blind single-state initialization initialized switching weights. Default configurations in Table~\ref{tab:configs} were used for other parameters. SDRs of observed signals were -4.7~dB (2-Mix) and -6.8~dB (3-Mix). }}\label{exp:2mix_sdr}
\end{figure}

\begin{table}[t]
    \centering
    \caption{{Separate evaluation of source separation (SIR improvement), denoising (noise power reduction ratio), and dereverberation (SDR improvement) with 3-Mix of TIMIT-ConvMix: Larger values indicate better performances with all the measures. }}
    \label{tab:derevdenoise}
    {
    \begin{tabular}{c|c|cc|c|cc}\toprule
         & IVA & \multicolumn{2}{c|}{swIVA} & CIVA & \multicolumn{2}{c}{swCIVA}\\\hline
    \rule[0mm]{0mm}{3mm}$I$ & - & - & - & 1 & 2 & 3\\
    $J$ & 1 & 2 & 3 & 1 & 2 & 3 \\\hline
    \rule[0mm]{0mm}{3mm}Separation (dB) & 4.70 & 5.29 & 5.53 & 7.11 & 7.90 & 8.08 \\
    Denoising (dB) & 6.54 & 6.64 & 6.65 & 6.91 & 7.12 & 7.19 \\
    Dereverberation (dB) & 1.24 & 1.59 & 1.96 & 3.44 & 3.77 & 3.92 \\\bottomrule
    \end{tabular}}
\end{table}
\begin{table}[t]
    \centering{
    \setlength{\tabcolsep}{2.2pt}
    \caption{Computing times in seconds required for processing mixtures in 2-Mix and 3-Mix of TIMIT-ConvMix with length of 13.1~s.}    \label{tab:ctime}
    \begin{tabular}{c|c|cc|c|cc|ccc|ccc}\toprule
         & {\IVA} & \multicolumn{2}{c|}{\swIVA} & {\IVAconv}& \multicolumn{8}{c}{\swIVAconv} \\\hline
        \rule[0mm]{0mm}{3mm}$I$ & - & \multicolumn{2}{c|}{-} & 1 & \multicolumn{2}{c|}{1} & \multicolumn{3}{c|}{2} & \multicolumn{3}{c}{3}\\
        $J$ & 1 & 2 & 3 & 1 & 2 & 3 & 1 & 2 & 3 & 1 & 2 & 3\\\hline
        \rule[0mm]{0mm}{3mm}2-Mix  & 5.6 & 8.5 & 10.9 & 11.5 & 17.1 & 20.7 & 20.8 & 29.0 & 37.3 & 28.9 & 40.6 & 47.2 \\
        \rule[0mm]{0mm}{3mm}3-Mix & 8.7 & 12.9 & 18.4 & 19.7 & 27.8 & 37.7 & 34.8 & 47.9 & 62.8 & 51.2 & 69.7 & 88.3 \\\bottomrule
    \end{tabular}
}
\end{table}
Figure~\ref{exp:2mix_sdr} shows the improvement in the SDRs obtained from the observed signals using {\IVA}, {\swIVA}, {\IVAconv}, and {\swIVAconv} after 50 updates of the separation matrices. Blind single-state initialization was used for {\swIVA} and {\swIVAconv}. The figure clearly demonstrates that SDRs consistently improved by respectively increasing $I$ and $J$ from $1$ to $3$ for both 2-Mix and 3-Mix cases. 

{The performance was also evaluated separately for source separation, denoising, and dereverberation using 3-Mix of the dataset. Given a filter estimated for each mixture, source separation was evaluated by signal-to-interference ratio (SIR) improvement \cite{BSSEVAL} obtained from the mixture. In contrast, denoising was evaluated by noise power reduction ratio obtained when we applied the filter to the noise in the mixture, and dereverberation was evaluated by SDR improvement obtained when we applied the filter to each (noiseless) reverberant speech signal in the mixture. Table~\ref{tab:derevdenoise} summarizes the results, where each measure exhibits the same improvement tendency as that in Fig.~\ref{exp:2mix_sdr}.}

Table~\ref{tab:ctime} shows the actual computing times required for processing mixtures whose length was 13.1~s in 2-Mix and 3-Mix. The processing code was implemented using Python version 3.7.10 with Numpy version 1.20.2 mainly to evaluate the estimation accuracy, but it was not necessarily optimized to evaluate the computing times. For example, the summation in Eqs.~(\ref{eq:Rupdate}) and (\ref{eq:qcov}) was calculated for all $t$ regardless whether $\beta_{t,f}^{(i,j)}$ takes a zero value, although it largely increases the computational complexity of Cov. Faster implementation could be enabled with lower level programming languages.

According to the table, the computing times increased as $I$ and $J$ increased, but the amount was less than the linear order on both $I$ and $J$. For example, for 2-Mix, the computing time increased as 11.5, 20.8, and 28.9 with $(I,J)=(1,1)$, $(2,1)$, and $(3,1)$, and further increased as 28.9, 40.6, and 47.2 with $(I,J)=(3,1)$, $(3,2)$, and $(3,3)$. 
In contrast, the number of microphones had a larger impact on the computing time. For example, the computing time increased from 47.2 to 88.3 when increasing the number of microphones from 3 (2-Mix) to 4 (3-Mix) for $(I,J)=(3,3)$.

\begin{figure}
\includegraphics[width=\columnwidth]{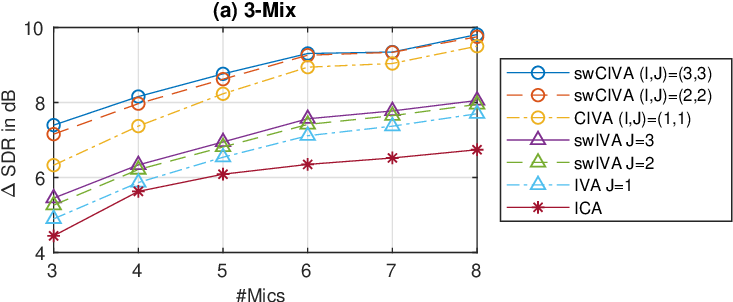}
\caption{{Improvement of SDRs in dB with varying number of microphones on 3-Mix of TIMIT-ConvMix using {\IVA}, {\swIVA}, {\IVAconv}, and  {\swIVAconv} after 50 updates of separation matrices: Performance of ICA is also shown as a baseline. Blind single-state initialization initialized switching weights. Default configurations in Table~\ref{tab:configs} were used for other parameters.}}\label{exp:mics}
\end{figure}

{Finally, Fig.~\ref{exp:mics} shows the SDR improvement on 3-Mix when we varied the number of microphones from 3 to 8. ICA is also included as a reference in addition to IVA, swIVA, CIVA, and swCIVA. ICA was optimized using IP \cite{IP} and followed by frequency permutation re-alignment \cite{sawada11aslp}. 
While all the methods improved the SDRs as the number of microphones increased, swIVA and swCIVA, respectively, outperformed IVA and CIVA under all the conditions.
}

\section{Concluding remarks}
\label{sec:conclude}
This paper proposed {\swIVA} and {\swIVAconv} that incorporated a switching mechanism into conventional {\IVA}-based source separation and {\IVAconv}-based joint source separation and dereverberation to improve their performance when only a relatively small number of microphones are available. The switching mechanism enables the improvement by clustering the time frames of an observed signal into groups that can be well handled by a small number of microphones and by assigning conventional {\IVA}/{\IVAconv} techniques to independently process individual groups. We introduced several essential techniques to let {\swIVA} and {\swIVAconv} work appropriately, including two initialization techniques based on blind and spatially guided approaches, a coarse-fine source model, a factorized switching model, and separation matrix-wise switching. Experiments showed that both {\swIVA} and {\swIVAconv} greatly outperformed the conventional {\IVA} and {\IVAconv} in terms of FWSSNRs, SDR, and ASR scores when using 2 or 3 microphones. In particular, we obtained consistent and substantial improvement when we respectively set the number of switching states at two for the MCLP filters and for separation matrices. 

{Future work may develop a more consistent source model that does not rely on the proposed hybrid model. For example, such advanced source models as non-negative matrix factorization \cite{ilrma2016} and neural networks \cite{Kitamura2019TASLP,Kameoka2019NC} might consistently solve such problems.  An evaluation with spatially guided initialization might also be important when the estimation accuracy of the spatial guide is degraded under more challenging adverse conditions.}

\appendices

\section{{Derivation of likelihood function}}\label{sec:derivobj}

{Assuming $\vect{x}_{t,f}$ is a zero vector for $t\le 0$, Eq.~(\ref{eq:tvCBF}) can be rewritten:
\begin{align}
{\bm y}_f&={\bm W}_f^{\HT}{\bm x}_f,\\
{\bm x}_f&=[\vect{x}_{T,f}^{\top},\vect{x}_{T-1,f}^{\top},...,\vect{x}_{1,f}^{\top}]^{\top}\in\mathbb{C}^{MT},\\
{\bm y}_f&=[\vect{y}_{T,f}^{\top},\vect{y}_{T-1,f}^{\top},...,\vect{y}_{1,f}^{\top}]^{\top}\in\mathbb{C}^{MT},\\
{\bm W}_f&=
\begin{bmatrix}
\vect{W}_{T, f} & O & & & O \\
\ast & \vect{W}_{T - 1, f} & \ddots & & \\
 & \ast & \ddots & O & \\
 && \ddots & \vect{W}_{2, f} & O \\
O &&& \ast & \vect{W}_{1, f} 
\end{bmatrix}
\end{align}
where ${\bm x}_f$ and ${\bm y}_f$ are vectors containing the whole time series of $\{\vect{x}_{t,f}\}_t$ and $\{\vect{y}_{t,f}\}_t$,  ${\bm W}_f\in\mathbb{C}^{MT\times MT}$ is a lower triangular block matrix containing the whole time series of $\{\vect{W}_{t,f}\}_t$ in its block diagonal components, and $\ast$ represents certain components corresponding to $\{\bar{\vect{W}}_{t,f}\}_t$.
According to the assumption in Eq.~(\ref{eq:independence}) and a property of a triangular block matrix, the log likelihood function can be expanded:
\begin{align}
    \hspace{-1.5mm}\log p\left(\{{\bm x}_f\}_f\right) &=\sum_{f=1}^F\log p\left({\bm W}_f^{-\HT}{\bm y}_f\right),\\
    &\hspace{-1.9cm} =\sum_{f=1}^F\log p({\bm y}_f) + 2\sum_{f=1}^F\log|\det {\bm W}_f|,\\
    &\hspace{-1.9cm}=\sum_{n=1}^M\sum_{t=1}^T\sum_{f=1}^F\log p({y}_{n,t,f})+2\sum_{t=1}^T\sum_{f=1}^F\log|\det\vect{W}_{t,f}|. 
\end{align}}

\footnotesize
\bibliographystyle{IEEEtran}
\bibliography{IEEEabrv,bibs}
\end{document}